\documentclass[%
 reprint,
superscriptaddress,
 amsmath,amssymb,
 aps,
]{revtex4-2}

\usepackage[
pagebackref=false,
colorlinks=true,
linkcolor=blue,
urlcolor=blue,
filecolor=black,
citecolor=red,
pdfstartview=FitV,
pdftitle={},
pdfauthor={},
pdfsubject={},
pdfkeywords={},
pdfpagemode=None,
bookmarksopen=true
]{hyperref}

\usepackage{graphicx}
\usepackage{dcolumn}
\usepackage{bm}
\usepackage{enumerate}
\usepackage{here}
\usepackage{xcolor}
\usepackage{bbm}
\usepackage{braket}

\newcommand{\ra}{\rightarrow}
\newcommand{\R}{\textcolor{red}}
\newcommand{\mS}{\mathcal{S}}

\newcommand{\tr}{\mathrm{tr}}
\newcommand{\bo}{\mathbbm{1}}

\newcommand{\dg}{\dagger}

\newcommand{\mr}{\mathrm}

\newcommand{\ex}[1]{\mathrm{e}^{#1}}

\newcommand{\pa}[1]{\left(#1 \right)}

\newcommand{\ca}[1]{\mathcal{#1}}

\newcommand{\fr}{\frac}

\def\tr{{\text{tr}}}

\begin{document}

\preprint{APS/123-QED}
\preprint{CALT-TH 2022-043}
\preprint{RIKEN-iTHEMS-Report-22}

\title{
  Multipartite entanglement in two-dimensional chiral topological liquids 
}

\date{\today}

\author{Yuhan Liu}
\affiliation{Kadanoff Center for Theoretical Physics, University of Chicago, Chicago, IL~60637, USA}
\affiliation{James Franck Institute, University of Chicago, Chicago, Illinois 60637, USA}

\author{Yuya Kusuki}
\affiliation{
Walter Burke Institute for Theoretical Physics,
California Institute of Technology, Pasadena, CA 91125, USA}
\affiliation{
RIKEN Interdisciplinary Theoretical and Mathematical Sciences (iTHEMS),
Wako, Saitama 351-0198, Japan}

\author{Jonah Kudler-Flam}
\affiliation{
School of Natural Sciences, Institute for Advanced Study, Princeton, NJ 08540 USA}
\affiliation{
Princeton Center for Theoretical Science, Princeton University, Princeton, NJ 08544, USA}

\author{Ramanjit Sohal}
    \affiliation{Department of Physics, Princeton University, Princeton, New Jersey, 08544, USA}

\author{Shinsei Ryu}
\affiliation{Department of Physics, Princeton University, Princeton, New Jersey, 08544, USA}

\begin{abstract}
The multipartite entanglement structure for the ground states of two dimensional topological phases is an interesting albeit
 not well understood question. Utilizing the bulk-boundary correspondence, the calculation of tripartite entanglement in 2d topological phases can be reduced to that of the \textit{vertex state}, defined by the boundary conditions at the interfaces between spatial regions. In this paper, we use the conformal interface technique to calculate entanglement measures in the vertex state, 
 which include area law terms, 
 corner contributions, and topological pieces,
 and a possible additional order one contribution. This explains our previous observation of the Markov gap $h=\frac{c}{3}\ln 2$ in the 3-vertex state, and generalizes this result to the $p$-vertex state, general rational conformal field theories, and more choices of subsystems. Finally, we support our prediction by numerical evidence, finding precise agreement. 
\end{abstract}

\maketitle


\section{Introduction}

Topologically ordered phases of matter are characterized not by local order
parameters but by their pattern of long-range entanglement.
Concretely,
from the scaling of the entanglement entropy of a
bipartition,
one can extract
a universal property of topological ground states,
the so-called topological entanglement entropy
\cite{2006PhRvL..96k0404K,2006PhRvL..96k0405L}.
While the topological entanglement entropy 
provides a signature for 
topological ground states,
it is far from a full characterization of topological data;
there are topological states that share the same topological entanglement
entropy yet are distinct from each other.

Recent publications
\cite{2022PhRvB.105k5107L,siva2022universal,2022PhRvL.128q6402K,2022PhRvB.106g5147K} initiated the study of multipartite entanglement of topological ground states in
two spatial dimensions.
In particular, Refs.\
\cite{2022PhRvB.105k5107L,siva2022universal} investigated the recently introduced reflected entropy
\cite{2021JHEP...03..178D}
as well as the entanglement negativity
for tripartitions of topological ground states, in which the three spatial subregions meet at junctions.
For a system of three spin 1/2 degrees of freedom,
the reflected entropy
(or more precisely, the Markov gap which 
is the difference between
the reflected entropy and mutual information 
-- see below)
detects the tripartite entanglement of the
W-state, 
while 
it is insensitive 
to GHZ-type 
tripartite entanglement
\cite{Akers_2020}. 
The reflected entropy can thus capture quantum correlations beyond simple Bell
or EPR-type bipartite correlations. Indeed, a structure theorem was proven in Ref.~\cite{2021PhRvL.126l0501Z}, classifying the set of states with zero Markov gap \footnote{See Ref.~\cite{2021JHEP...10..047H} for further discussion that gave the Markov gap its name.}. 
There, the reflected entropy
was further studied
for gapped as well as critical ground states in one spatial dimension.

Ref.\ \cite{2022PhRvB.105k5107L}
discussed the reflected entropy $S_R(A,B)$ for two regions $A$ and $B$
for the integer quantum Hall ground state
and
the ground state of a two-dimensional chiral $p$-wave superconductor.
Here, the total system is put on a spatial sphere and tripartitioned
into three regions $A$, $B$, and $C$ that meet at two points (junctions). 
The region $C=\overline{AB}$ is partially traced out to obtain the reduced density
matrix for $A\cup B$.
(For the definition of reflected entropy, see Sec.\ \ref{sec:reflected-entropy}.)
By
using the bulk-boundary correspondence and techniques from string field theory,
Ref.\ \cite{2022PhRvB.105k5107L}
found that the Markov gap, which is
the difference between the reflected entropy and the mutual information,
is independent of the subregion sizes and given by the universal formula
\begin{align}
  \label{h and c}
h(A,B) :=S_R(A,B) -I(A,B) = \frac{c}{3}\ln 2,
\end{align}
where $c$ is the central charge of the topological liquid, and
$c=1$ and $c=1/2$ for the integer quantum Hall state and chiral $p$-wave
superconductor, respectively.
On the other hand, Ref.\ \cite{siva2022universal} studied
the Markov gap for string-net models (the Levin-Wen models),
for which the central charge is zero, 
and found that $h(A,B)=0$. 
(See also Ref.\ \cite{Sohal22}.)
Equation \eqref{h and c}
relates tripartite quantum entanglement to the central charge and hence
captures the universal data of topological liquid beyond the topological
entanglement entropy.
These calculations were done for ideal, representative
topological ground states realized deep inside a topological phase.
Refs.\ \cite{2022PhRvB.105k5107L, siva2022universal}
also numerically studied the reflected entropy in
a lattice model of Chern insulators.
While the Markov gap was still found to be insensitive to the subregion and systems
sizes,
the formula \eqref{h and c} does not hold verbatim,
but its RHS
provides a lower bound of the Markov gap,
$h(A,B) \ge (c/3) \ln 2$.
Putting these results together,
the Markov gap in the tripartition setup above is conjectured to
capture the central charge of stable (ungappable) degrees of freedom at the
boundary of topological liquid;
that is to say, a non-zero Markov gap may be an obstruction to completely gap out the boundary (edge) theory.

Despite these recent results,
Eq. \eqref{h and c} has been verified only for a fraction
of topological liquids -- the Levin-Wen models for non-chiral topological order, 
and the free fermion models (integer quantum Hall and topological superconductor
states). Thus, the majority of interacting chiral topological states
with non-zero chiral central charge have not been discussed. 
Also, the free fermion results are half-numerical;
while
Ref.\ \cite{2022PhRvB.105k5107L}
analytically constructed the vertex state -- an essential ingredient for
achieving multipartition in our approach
(see Sec.\ \ref{sec:Corner contributions and vertex states} for details), 
the relevant entanglement quantities (reflected entropy, Markov gap,
entanglement negativity, etc.) had to be computed numerically.  
Hence, an analytic understanding of the Eq. \eqref{h and c} has been lacking.

In this work, we present an analytical approach
to multipartite entanglement that can be applied to
generic (chiral) topological ground states in two spatial
dimensions. 
Following Ref.\ \cite{2022PhRvB.105k5107L},
we use the bulk-boundary correspondence and reduce the
calculations of the entanglement quantities 
to calculations in conformal field theory (CFT).
Specifically,
we will show that  
the calculations can be reformulated in terms of defect (interface) CFT,
and furthermore simplified
by taking the advantage the limit of small $\beta/L$, where $1/\beta$
is the bulk gap and $L$ is the length of the entangling boundary.  
By using a series of conformal transformations,
a given entanglement quantity can be evaluated as a path integral
on a cylinder with topological interfaces.

In addition to reflected entropy and the Markov gap,
this approach also allows us to calculate the so-called corner contribution 
to the bipartite entanglement entropy studied
in Refs.\ 
\cite{Rodr_guez_2010,2021PhRvB.103k5115S,Ye_2022}.
In these works,
the subregions for bipartitions containing sharp corners or cusps
were considered. 
It was found that
the entanglement entropy 
receives a geometric angle-dependent contribution. 
(See also 
several works on closely-related 
quantities and setups, 
such as the charge fluctuations
for a subregion with 
a sharp corner
\cite{Estienne_2022,
https://doi.org/10.48550/arxiv.2211.05159},
and others
\cite{Estienne_2020,
10.3389/fphy.2022.971423,
https://doi.org/10.48550/arxiv.2208.12819}.
Quantum Hall states
on surfaces with cusp singularities
were also studied
in the literature
\cite{1992PhRvL..69..128A, Can_2017, 2017JPhA...50W4003C}.
As we will see,
the corner contribution 
from our approach,
Eq.\ \eqref{eqn:ent-p-vertex},
does not seem to 
match precisely  
with the previous works above.
We will speculate on the possible source of 
the discrepancies. 
Nevertheless, we will show that 
our prediction \eqref{eqn:ent-p-vertex}
agrees with numerically-computed 
entanglement quantities 
for four vertex states in the free fermion theory. 
Here, we extend
the construction of
three-vertex states
(vertex states that can be used to tripartition a topological liquid)
in the free fermion theory 
in Ref.\ \cite{2022PhRvB.105k5107L}
to four vertex states.
This allows us to tetrapartition the topological liquid,
and test our analytical predictions for the 
corner contribution
and the reflected entropy (Markov gap).

The rest of the paper is organized as follows.
In Sec.\ \ref{sec:cft},
we revisit the calculation of
bipartite entanglement entropy for topological ground states in two dimensions
using the bulk-boundary correspondence. 
As is well known, the calculation can be formulated in terms of
boundary states in boundary conformal field theory (BCFT). 
For later use,
we reformulate the calculation in terms of defects (interfaces)
in CFT and also show that the calculation can be simplified
in the $\beta/L \to 0$ limit.
In 
Sec.\ \ref{sec:Corner contributions and vertex states},
we generalize and extend the approach of  
Sec.\ \ref{sec:cft} to multipartitions
by considering $p$-vertex states ($p>2$).
This allows us to calculate
the 
corner contribution to bipartite entanglement entropy
and
the reflected entropy.
In Sec.\ \ref{sec:numerics}, we present the construction of four-vertex states
and the numerical calculations of entanglement quantities.
We conclude with a discussion in Sec.\ \ref{sec:disc}.

\section{Edge theory approach: Review of Bipartite Entanglement Entropy}
\label{sec:cft}

As noted in the Introduction, our methodology for computing entanglement in chiral topological orders employs the boundary state or ``cut-and-glue" approach. This method reduces computations of bulk entanglement quantities to computations of the same quantities in the corresponding edge CFT, for which powerful analytic techniques exist. Indeed, a central technical advance of our work is the use of defect CFT methods to extend the cut-and-glue approach to the computation of 
the corner contributions
as well as multipartite entanglement quantities. 
In this section, as a warm-up, we review this approach and introduce our defect CFT formalism in the simpler setting of the computation of the bipartite entanglement entropy, for which the result in a chiral topological order is well known.

To that end, let us consider a chiral topological order on the surface of a sphere. We are interested in computing the entanglement entropy for a spatial bipartition of the sphere into two regions $A$ and $\bar{A}$, which we take to be the two hemispheres, such that the entanglement cut lies along the equator. In the cut-and-glue approach to this problem, we \emph{physically} cut the system along the entanglement cut, giving rise to counter-propagating chiral CFTs along the new edges. We can then ``heal" the cut by introducing appropriate tunnelling terms to gap out the CFTs. Now, since the correlation length is effectively zero in the bulk, we may approximate the entanglement between the bulk regions $A$ and $\bar{A}$ as arising purely from degrees of freedom near the entanglement cut, namely the gapped edge degrees of freedom. The upshot of the cut-and-glue approach is the reduction of the problem to a computation of the entanglement entropy between the left and right movers in the gapped interface. This approach naturally generalizes to the primary interest of this work, namely a partitioning of a topological phase into three or more spatial subregions, meeting at two junctions. Computations of multipartite entanglement quantities then reduce to computations of the same quantities in a network of multiple gapped interfaces meeting at two junctions, as we will discuss in more detail in Sec.\ \ref{sec:Corner contributions and vertex states}.

In the remainder of this section, we spell out the computation in the bipartite setting in more detail, with an eye towards setting the stage for the more involved multipartite computations. In particular,
for the bipartite case at hand, it is well-established that an appropriate approximation to the ground state of the gapped interface is provided by conformal boundary states known as Ishibashi states \cite{Ishibashi:1988kg,qi2012general,2016PhRvB..93x5140W}. 

As we shall review below in Sec. \ref{sec:Boundary states and left-right entanglement entropy}, the explicit form of the Ishibashi states are known for generic 
rational CFTs,
allowing for a simple and direct computation of the bipartite entanglement entropy, reproducing the known result for the topological entanglement entropy. 
Similarly, for a multipartitioning, as described in Ref.\ \cite{2022PhRvB.105k5107L} and reviewed below in Sec. \ref{sec:Corner contributions and vertex states}, the configuration of gapped interfaces can be approximated by a so-called \emph{vertex} state. The explicit form of such states for generic rational CFTs is \emph{not} known, necessitating an alternative approach to computing the entanglement. With this in mind, in Sec.\
\ref{sec:conformal-interface-approach}, we introduce a complementary path integral approach to computing the bipartite entanglement entropy, making use of conformal interfaces. Finally, in Sec. \ref{sec:path-int-dec}, we present an approximation of the preceding path integral, which reduces the computation of the entanglement of a CFT partition function on a torus to one of CFT partition functions on two cylinders. We emphasize that these latter two subsections, while simply reproducing known results for the bipartite entanglement entropy, lay the technical foundations for the subsequent computations of the 
corner contributions to the bipartite entanglement entropy in Sec.\ \ref{sec:Corner contributions and vertex states} and the reflected entropy in Sec.\ \ref{sec:reflected-entropy}.


\subsection{Boundary states and left-right entanglement entropy}
\label{sec:Boundary states and left-right entanglement entropy}

We begin by first reviewing in more detail the CFT description of the interface and the computation of the bipartite entanglement entropy. As noted above, after ``gluing" the edges to heal the physical cut along the entanglement cut, the degrees of freedom along the cut may be described by an Ishibashi state. 

Formally, Ishibashi states are states satisfying conformal boundary conditions in a non-chiral CFT. 
Explicitly, let us consider a non-chiral CFT defined
on a spatial circle of length $L$.
A boundary state $|B\rangle$
is defined to satisfy 
\begin{equation}
    [T(\sigma)-\bar{T}(\sigma)]|B\rangle=0,
    \label{eqn:boundary-condition-B}
\end{equation}
where
$T(\sigma)$ and $\bar{T}(\sigma)$
are the holomorphic and antiholomorphic components of
the stress-energy tensor, and
$\sigma$ is the spatial coordinate
\cite{0411189}.
An Ishibashi state 
satisfies, in addition to 
\eqref{eqn:boundary-condition-B},
similar conditions for conserved currents.

Now, the CFTs describing the edges of chiral topological orders are rational. That is to say, the Hilbert space can be decomposed into a finite number of primary operator sectors labeled by $i$: $\mathcal{H}=\oplus_i (\mathcal{V}_i\otimes \bar{\mathcal{V}}_i)$. The bulk-boundary correspondence states that the anyon $i$ in the bulk is associated with a primary operator $i$ in the boundary CFT. The chiral sector $\mathcal{V}_i$ has orthonormal basis $|h_i,N,k\rangle$, where $h_i$ is the conformal dimension of the primary operator $i$, $N$ is the level that goes from $0$ to $\infty$, and $k$ labels the  $d_{h_i}(N)$ degenerate states in a given level $N$. Similarly, the anti-chiral sector $\bar{\mathcal{V}}_i$ has orthonormal basis $\overline{|h_i,N,k\rangle}$, which is isomorphic to $\mathcal{V}_i$. For these rational CFTs, solutions to Eq. \eqref{eqn:boundary-condition-B} are provided by the Ishibashi states, which take the form,
\begin{equation}
    |B_i\rangle=\sum_{N=0}^\infty \sum_{k=1}^{d_{h_i}(N)}|h_i,N;k\rangle\otimes \overline{|h_i,N;k\rangle}.
\end{equation}
There is a single Ishibashi state for each primary, or anyon, sector $i$. General solutions to the conformal boundary conditions are given by linear combinations of the Ishibashi states. 
Now, the Ishibashi states $|B_i\rangle$ are not normalizable. In order to define a physical state, we introduce a regulator $\beta$ and construct the regularized state $|\mathbf{B}_i\rangle$:
\begin{equation}
|B_i\rangle\ra |\mathbf{B}_i\rangle= \frac{e^{-\beta H_0}}{\sqrt{n_i}}|B_i\rangle.
\end{equation}
Here $H_0$ is the CFT Hamiltonian $H_0=\frac{2\pi}{L}(L_0+\bar{L}_0-\frac{c}{12})$, where $L_0,\bar{L}_0$ are zero modes of the stress tensor $T,\bar{T}$, $c$ is the central charge, and $L$ is the length of the entanglement cut. The normalization factor $n_i$ is defined such that these regularized states are orthonormal $\langle \mathbf{B}_i|\mathbf{B}_j\rangle=\delta_{ij}$. 

Returning to our problem of computing the entanglement entropy, Ishibashi states were argued to describe the gapped interface along the entanglement cut in Ref. \cite{qi2012general}. Indeed, the conformal boundary condition (and the explicit form of the Ishibashi states) pairs up the left and right movers, just as a tunneling interaction would gap out the left and right movers at the interface. The regulator $\beta$, in this description, is determined by the inverse bulk gap, and hence we will always be interested in the $\beta \to 0$ limit. In the context of the bipartition of the sphere, the Ishibashi state $\ket{\textbf{B}_i}$ thus describes the gapped interface along the entanglement cut when a single anyon line $i$ pierces the entanglement cut.

With the description of the interface in hand, we can approximate the entanglement entropy between the bulk regions $A$ and $\bar{A}$ as the ``left-right" entanglement entropy of the boundary state -- that is, the entanglement entropy between the chiral and anti-chiral sectors of the interface Hilbert space. Explicitly, for an interface state of the form $
\ket{\psi} = \sum_i c_i \ket{\textbf{B}_i}$, corresponding to the density matrix
\begin{equation}
  \label{initial density matrix}
  \rho =\sum_{i,j} \frac{c_i c_j^*}{\sqrt{n_i n_j}} e^{-\beta H_{0}}\ket{B_i}\bra{B_j} e^{-\beta H_{0}}.
\end{equation}
describing a superposition of states with single anyon lines threading the entanglement cut,
we can explicitly compute the reduced density matrix for the left-movers (or chiral sector),
\begin{align}
    \rho_L = \mathrm{Tr}_R
    \big[\ket{\psi}\bra{\psi}\big],
\end{align}
where the trace is performed over the right-movers (anti-chiral sector). The entanglement entropy is then straightforwardly computed using the replica trick,
\begin{align}
    S_A = -\mathrm{Tr}\, 
    \big[\rho_L \log \rho_L \big] = \lim_{n \to 1} \frac{1}{1-n} \ln \mathrm{Tr}\, \rho_L^n.
\end{align}
Carrying this through, one finds
\begin{equation}
  \label{top ent}
    S_{A} = \frac{c\pi L}{24\beta} - S_{\mathrm{topo}},
\end{equation}
where the first is the non-universal area law contribution
while the second term is the topological entanglement entropy,
\begin{equation}
\begin{aligned}
S_{\mathrm{topo}} & =\ln\mathcal{D}+\sum_i |c_i|^2\ln |c_i|^2-\sum_i|c_i|^2 \ln d_i,
    \label{eq:topo-ent}
\end{aligned}
\end{equation}
where $d_i$ is the quantum dimension of anyon $i$ and $\mathcal{D} = \sqrt{\sum_i d_i^2} $ is the total quantum dimension. 
For an Ishibashi state $c_i=\delta_{i,a}$, corresponding to a state with a definite anyon line threading the entanglement cut, the topological contribution is $S_{\mathrm{topo}}=\ln\mathcal{D}-\ln d_a$.
For Abelian topological order, all quantum dimensions $d_i=1$
and the topological entanglement entropy reduces to  $S_{\mathrm{topo}}=\ln
\mathcal{D}+\sum_i|c_i|^2 \ln |c_i|^2$, where the second term is
the Shannon entropy associated with
the probability distribution $\{|c_i|^2\}$. We thus see that the boundary state description of the interface captures the expected bipartite entanglement structure of the bulk topological phase.

For the convenience of later discussion, let us also mention a more complicated case of two anyon lines insertion $a$ and $b$. The coefficient $c_i$ can be computed as \cite{2016PhRvB..93x5140W}: $|c_i|^2=N_{ab}^i \frac{d_i}{d_a d_b}$ where $N_{ab}^i$ is the fusion coefficient. One can show the topological contribution is  $S_{\mathrm{topo}}=\ln (\mathcal{D}/d_a d_b)$ in the case $N_{ab}^i=0$ or 1. 




\subsection{Conformal interface approach} \label{sec:conformal-interface-approach}

As described at the beginning of this section, a direct computation of the entanglement entropy from the ground state will not be possible when we later consider multipartitions of the bulk topological phase. While the ground state will satisfy similar conformal boundary conditions as the above Ishibashi states, its explicit form for general CFTs is not known. With this in mind, we now develop an alternative approach to computing the bipartite entanglement entropy using path integral and conformal interface methods \cite{2008JHEP...12..001S, 2016ForPh..64..516B}, which \emph{will} generalize to the multipartite case. Our strategy will be to re-express the R\'enyi moments of (reduced) density matrices formed from regularized boundary states as CFT partition functions on closed manifolds with insertions of conformal interfaces through a series of conformal mappings. 

\begin{figure}[t]
    \centering
    \includegraphics[width=\linewidth]{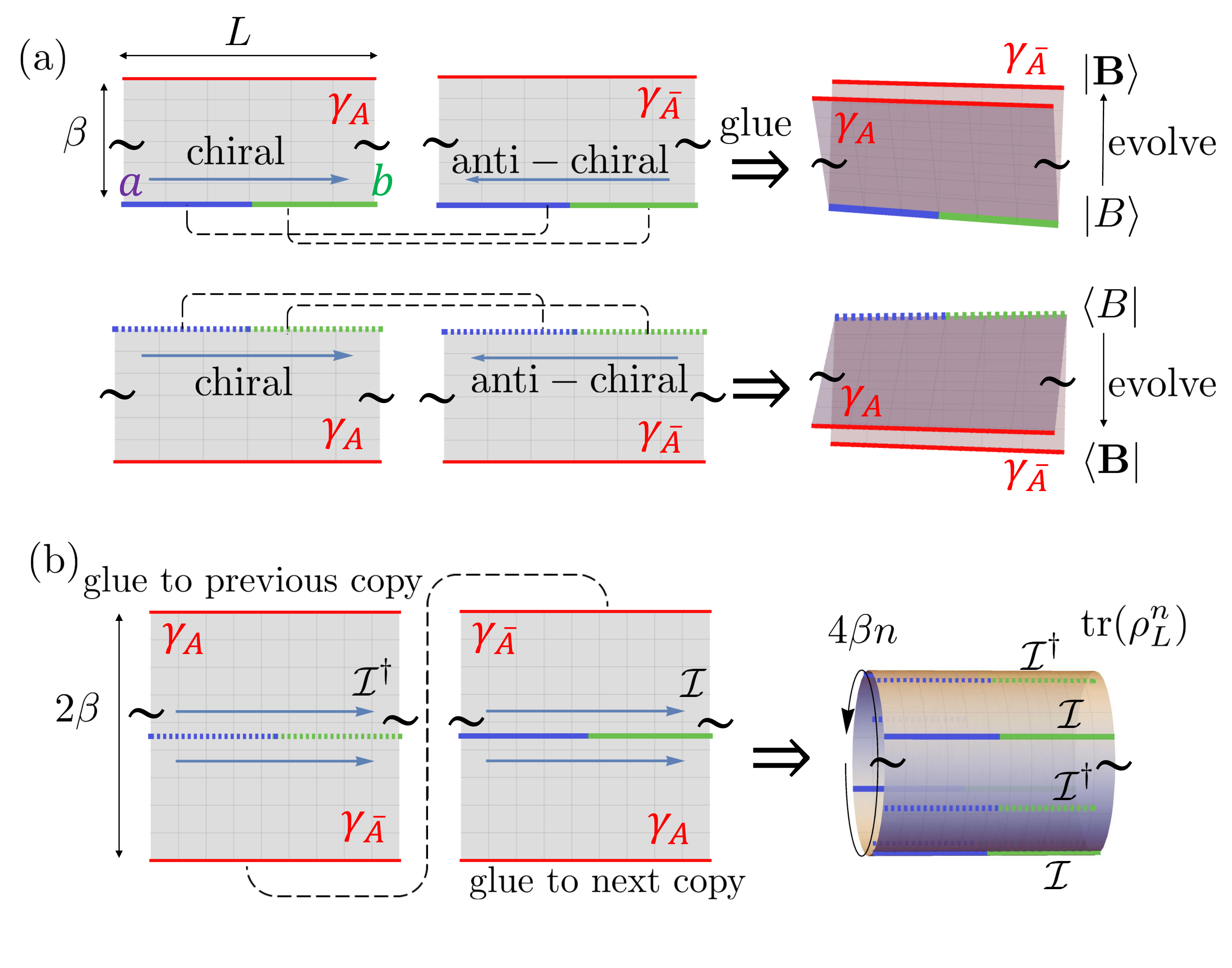}
    \caption{
      (a) Spacetime path integral representation
      of the regularized boundary state $|\mathbf{B}\rangle$
      starting from the unregularized boundary state $|B\rangle$
      as the initial condition,
      as well as $\langle B|$ and $\langle \mathbf{B}|$ as path integral. The vertical direction denotes the imaginary time evolution. ``gluing'' amounts to identifying chiral and anti-chiral copies, and we consider the case of two anyon lines insertion $a$ and $b$.  The black tilde symbol means taking periodic boundary condition. 
      (b) Unfolding the boundary state. In this picture $Z_n=\tr(\rho_L^n)$ can be expressed as path integral of a chiral theory on the torus with interfaces inserted. 
    }
    \label{fig:boundary_s}
\end{figure}

First, we start by expressing the density matrix $\rho$ 
 in Eq.\ \eqref{initial density matrix} as a path integral as depicted pictorially in Fig.\ \ref{fig:boundary_s}(a). 
The top and bottom rows correspond, respectively, to the ket $\ket{\textbf{B}_i}$ and bra $\bra{\textbf{B}_i}$. Focusing first on the ket, the two spacetime sheets correspond to the chiral and anti-chiral sectors of the the theory. The horizontal direction is the spatial direction, of length $L$, with the tildes indicating periodic boundary conditions. The bottom of the path integral denotes the unregularized boundary state $\ket{B_i}$. The dashed lines connecting the segments $a$ and $b$ indicate how the chiral and anti-chiral sectors are glued together as per the conformal boundary condition. The vertical direction denotes imaginary time evolution by $\beta$, yielding the regularized state $\ket{\textbf{B}_i}$.
More generally, 
we can start from 
a linear superposition $\sum_i c_i \sqrt{n_i}^{-1} | B_i\rangle$,
and the Euclidean path integral 
prepares the regularized state
$\sum_i c_i \sqrt{n_i}^{-1} e^{-\beta H_0} | B_i\rangle$.
%

Combining these elements yields the glued path integral representation of the ket in the right-most column.
Similar considerations hold for the bra, $\bra{\textbf{B}_i}$, represented in the second row of the same subfigure. Combining the bra and ket gives the path integral representation of the density matrix $\rho = \ket{\textbf{B}_i} \bra{\textbf{B}_i}$. 
When computing the trace $\tr(\rho)$, the top red lines of bra and ket are glued together ($\gamma_A$ to $\gamma_A$, $\gamma_{\bar{A}}$ to $\gamma_{\bar{A}}$), and the spacetime manifold becomes a torus.

Now, the computation of entanglement entropy requires computing the R\'enyi moments of the reduced density matrix, $\tr\, (\rho_L)^n$. These quantities can be represented as path integrals with conformal interface insertions, as we now demonstrate. We first take the path integral of $\rho$, and unfold the boundary state $|\mathbf{B}_i\rangle$ and $\langle\mathbf{B}_i|$, as shown in the left of Fig.\ \ref{fig:boundary_s}(b). After the unfolding, the doubled sheet of width $\beta$ becomes a single sheet with width $2\beta$. We define the interface operator $\mathcal{I}_i$ as the unfolded boundary state $|B_i\rangle$:
\begin{equation}
    \mathcal{I}_i = \sum_{N=0}^\infty \sum_{k=1}^{d_{h_i}(N)}|h_i,N;k\rangle\langle h_i, N;k|,
\end{equation}
where the ket of anti-chiral mode is ``flipped'' to the bra of chiral mode.  
In other words, $\mathcal{I}_i$ is an Ishibashi-type projector on the chiral sector $i$, $\mathcal{I}_i:\mathcal{V}_i\ra\mathcal{V}_i$. Note this is a well-defined projector since $|h_i,N;k\rangle$ is an orthonormal basis in $\mathcal{V}_i$. 
In general we can consider the linear superposition $\mathcal{I}=\sum_i c_i\sqrt{n_i}^{-1}\mathcal{I}_i$. 
Moreover, in the present context, these interfaces are topological, meaning they can be freely deformed in spacetime. 
Thus, after unfolding, each boundary state $\ket{B}$ ($\bra{B}$) maps to an insertion of the interface operator $\mathcal{I}$ ($\mathcal{I}^\dagger$).

\begin{figure}[t]
    \centering
    \includegraphics[width=\linewidth]{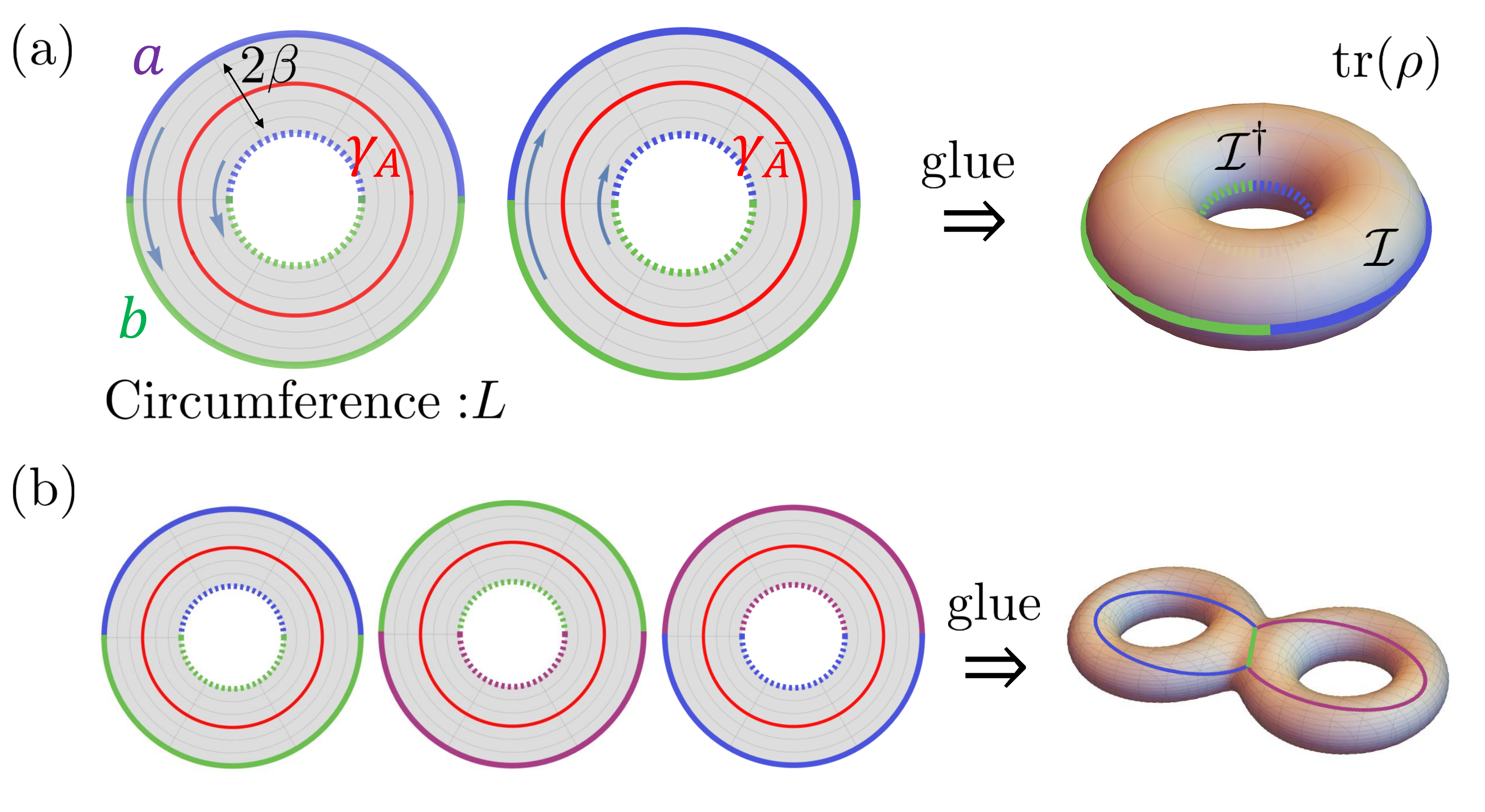}
    \caption{
      The spacetime path integral can be constructed by
      gluing annulus amplitudes (the annulus has width $2\beta$ and circumference $L$).
      Here, as an example, 
      $Z_1 = \tr \, (\rho)$ is constructed 
      for (a) $p=2$ and (b) $p=3$. After the gluing, the spacetime manifold becomes torus and 2-genus, respectively. 
    }
    \label{fig:boundary_s1}
\end{figure}

To obtain $\rho_L$, we take the path integral of $\mathcal{I},\mathcal{I}^\dg$, and glue the red lines $\gamma_{\bar{A}}$ for $\mathcal{I}$ and $\mathcal{I}^\dg$, as shown in the left of Fig.\ \ref{fig:boundary_s}(b) by the black dashed line. Then to obtain $\tr(\rho_L^n)$, we take $n$ copies of path integral of $\rho_L$, and glue the red line $\gamma_A$ for $\mathcal{I}$ of copy $i$ to the the red line $\gamma_A$ for $\mathcal{I}^\dg$ of copy $i+1$, where $i$ runs from $1$ to $n$. After this gluing, we obtain a torus with circumferences $L$ and $4\beta n$, and the path integral on this torus is $\tr\, (\rho_L^n)$:
\begin{equation}
Z_n\equiv \tr\, (\rho_L^n)=\tr\left((\mathcal{I}\mathcal{I}^\dg)^n e^{-4n\beta H_L}\right).
\end{equation}
The Hamiltonian $H_L$ in this expression is the \textit{chiral} Hamiltonian $H_L=\frac{2\pi}{L}(L_0-\frac{c}{24})$ \footnote{We treat our BCFT as an interface CFT (ICFT) by the unfolding. 
Then, the ICFT corresponds to a chiral part of the BCFT.}. This is shown in the right of Fig.\ \ref{fig:boundary_s}(b). 
In Fig.\ \ref{fig:boundary_s1}(a),
we show an equivalent path integral representation of $\tr(\rho_L)$ where the order of gluing is changed (first glue $\gamma_A,\gamma_{\bar{A}}$, then glue $a$ and $b$). In this way, the spacetime path integral can be constructed
by gluing annulus amplitudes where each annulus has width $2\beta$ and circumference $L$. The closed boundary condition is made explicit in this picture.

The path integral on the torus can be evaluated readily using standard CFT techniques. In CFT, the character of primary field sector $i$ is defined as $\chi_i(q)=\tr_{\mathcal{V}_i}(q^{L_0-\frac{c}{24}})$. As a first example, let's consider the special case of $Z_1$ with $c_i=\delta_{i,a}$. $Z_1$ can be directly related to the character:
\begin{equation}
Z_1=\tr\, (\mathcal{I}\mathcal{I}^\dg e^{-4\beta H_L})=\frac{1}{n_a}\chi_a(q),\quad q=e^{2\pi i(4\beta i/L)}.
\end{equation}
Since $Z_1=1$ by definition, one can read out the normalization factor $n_a=\chi_a(q)$. 

We now compute $Z_n$ in the large gap limit of interest $\beta\ra 0$. To work in this limit, we need to use the modular $\mathcal{S}$ transformation that brings $q=e^{2\pi i(4\beta i/L)}$ to $\tilde{q}=e^{2\pi i(-L/4\beta i)}$, and $\beta\ra 0$ corresponds to the limit of $\tilde{q}\ra 0$. The characters are related by $\chi_i(q)=\mS_{ii'}\chi_{i'}(\tilde{q})$ under the modular $\mathcal{S}$ transformation. Taking $\tilde{q}$ as a small quantity, the new character can be expanded as $\chi_{i'}(\tilde{q})=\tilde{q}^{h_{i'}-\frac{c}{24}}+\cdots$ and only the lowest order term shall be kept. By keeping only the lowest order term, $Z_n$ is: 
\begin{equation}\label{eq:Zn}
\begin{aligned}
    Z_n&=\tr\left(
   ( \mathcal{I} \mathcal{I}^\dg)^n e^{-4n\beta H_L} 
    \right)\\
    & = \sum_i \frac{|c_i|^{2n}}{\chi^n_i(q)} \chi_i(q^n)\\
    & = \sum_i |c_i|^{2n}\frac{\sum_{i'}\mS_{ii'}\chi_{i'}(\tilde{q}^{\frac{1}{n}})}{\left(\sum_{i'}\mS_{ii'}\chi_{i'}(\tilde{q})\right)^n}\\
    &\ra e^{\frac{\pi L c}{48\beta}(\frac{1}{n}-n)}\sum_i |c_i|^{2n} \mS_{i0}^{1-n},
\end{aligned}
\end{equation}
and we make another approximation to keep only the lowest order term in $\chi_0$: $\tilde{q}^{-c/24}$ 
\footnote{
While the first line appears to indicate 
that we have a single path-integral representation 
for $Z_n$, 
due to the normalization of the Ishibashi states, 
we need separate path integrals for 
the numerators and denominators. 
If we were to interpret $Z_n$ as a single path integral, 
the normalization factors would contribute 
to the spectrum 
as states with negative dimension.
This subtlety, however,
does not matter in the $\beta/L\to 0$ limit
where we only keep the vacuum block
as in the last line.
}.
Using this result, the entanglement entropy between the bulk
regions $A$ and 
 $\bar{A}$ is:
\begin{equation}
\begin{aligned}
    S_A&=\lim_{n\ra 1}\frac{1}{1-n}\ln Z_n\\
    &=\frac{\pi L c}{24\beta}-\sum_i |c_i|^2\ln |c_i|^2+\sum_i |c_i|^2\ln \mS_{i0}.
\end{aligned}
\end{equation}
Recalling $\mS_{i0}=d_i/\mathcal{D}$, 
this reproduces the known result of Eq.\ \eqref{top ent} 
 and Eq.\ \eqref{eq:topo-ent}.
We stress that although, at this point, conformal interface is simply a
reformulation of boundary CFT approach,
the path integral viewpoint will be useful
in the later discussion of tripartite entanglement. 
Finally,
we also note that 
for the pure state density matrix $\rho$, the
reflected entropy (to be defined in  Sec.\ \ref{sec:reflected-entropy}) is simply related to the entanglement entropy $S_A$ by
$S_{R}=2S_A$.
One can verify this relation using
conformal interface approach.
We leave the details to Appendix \ref{app:ref-pure}.

\subsection{Path integral decomposition}
\label{sec:path-int-dec}

Before concluding our discussion of bipartite entanglement, we introduce one final ingredient. The preceding path integral computation was tractable, as it amounted to computing CFT partition functions on the torus. When we turn to multipartite configurations, the exact path integral representation of the entanglement quantities of interest will be defined on higher genus surfaces, for which the partition functions are not readily obtained -- see Fig.\ \ref{fig:boundary_s1}(b) for an example of the tripartite configuration. Here, we make use of a pair-of-pants decomposition, valid in the $\beta \to 0$ 
(i.e., 
large bulk gap) limit, which reduces the torus path integral to a product of two cylinder path integrals. This decomposition will likewise simplify the multipartite path integrals to render the computations tractable.


 \begin{figure}[t]
 \begin{center}
  \includegraphics[width=8.0cm,clip]{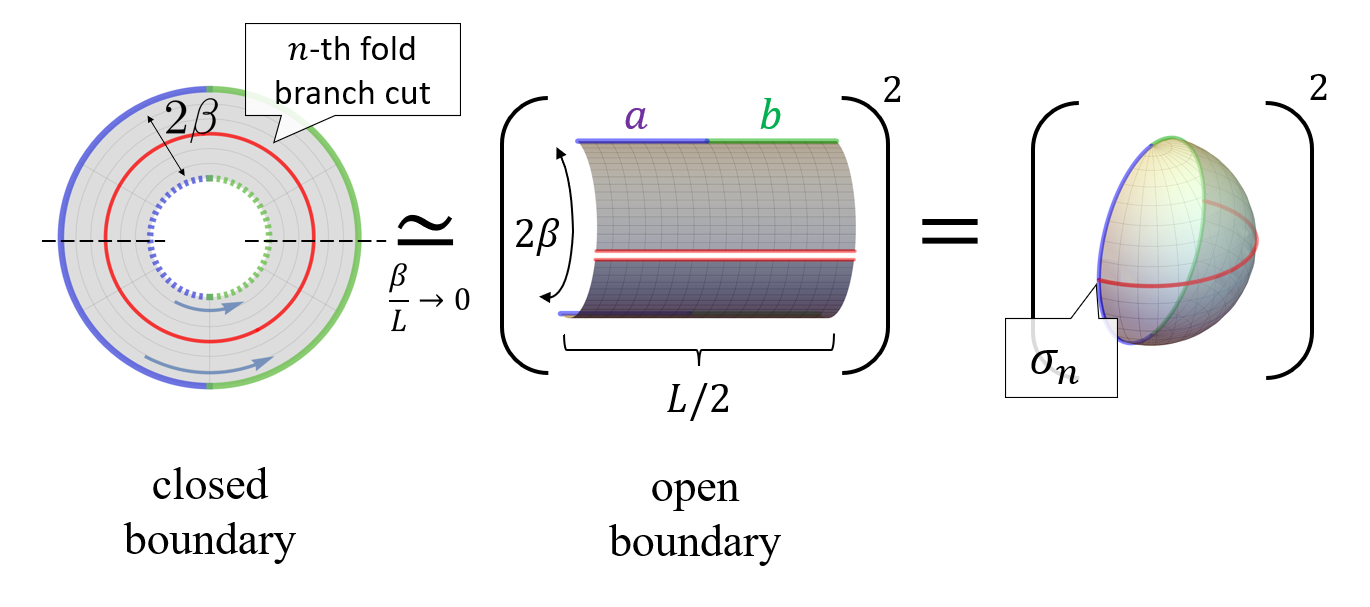}
 \end{center}
 \caption{
   Pants-like decomposition of $Z_{\text{closed}}$.
 In the limit ${\beta}/{L} \to 0$,  the replica partition function $Z_{\text{closed}}$ can be approximated by  $(Z_{\text{open}})^2$ (Center Figure). Each of the two strips has width $2\beta$ and length $L/2$. 
 The partition function $Z_{\text{open}}$ can be thought of as a disk partition function with two twist operators (Right Figure).
}
 \label{fig:app}
\end{figure}

We first note that the replicated path integral (partition function)
$Z_n = \mathrm{Tr}\, \rho_L^n$
can be obtained by gluing the annulus in Fig.\ \ref{fig:boundary_s1}(a).
In the limit $\beta/L \to 0$, 
the annulus is thin and hence only
the lowest dimension states
(lowest energy states)
propagate, taking the spatial direction as a fictitious time direction. 

Consider 
cutting the annulus into two strips with length $L/2$ and width $2\beta$
(Fig.\ \ref{fig:app}). 
Here, cutting the strip is equivalent to inserting a complete set of states at the cut.
When crossing the cut, the gluing condition is changed so the leading non-zero contribution in the complete set comes from:
\begin{equation}
  \sum_p \ca{W}_p \frac{|p\rangle\langle p|}{\langle p|p\rangle} \simeq  \ca{W}_{p_0} \frac{|p_0  \rangle\langle  p_0 |}{\langle p_0 | p_0 \rangle},
\end{equation}
where $p_0$ are the lowest energy states, which correspond to the twist
operator $\sigma_n$ with conformal dimension $h_{p_0}=h_{\sigma_n} \equiv
\fr{c}{24}\pa{n-\fr{1}{n}}$. The (regularized) normalization factor is
\cite{2022arXiv220604630K}:
\begin{equation}
  \mathcal{N}_n=\langle p_0 | p_0  \rangle=\sum_i |c_i|^{2n}
  \mathcal{S}_{i0}^{1-n}.
     \label{eq:norm-s}
 \end{equation}
 In general, we interpret the constant $\ca{W}_p$ as the number of states $p$.
 However, in this case, this is not necessary to be an integer
 because the normalization of the Ishibashi state depends on the moduli parameter.
We can show that
\begin{equation}\label{eq:dof}
 \ca{W}_{p_0} = \ca{N}_{n}.
\end{equation}
This just comes from the closed 
channel expansion of (\ref{eq:Zn}),
\begin{equation}
Z_n \simeq  \ca{W}_{p_0} \ex{- h_{p_0} \fr{\pi L}{2\beta}}.
\end{equation}
Before giving the expression in terms of these constants,
we would first like to explain the motivation for expressing the partition function in the closed string channel expansion (i.e. the quantization with the time direction along the interface).
Unfortunately, it is difficult to fix these theory-dependent constants [e.g. the normalization factor (\ref{eq:norm-s}) and the coefficients $\ca{W}_p$ (\ref{eq:dof})] in general, as we will see later in Sec.\ \ref{sec:reflected-entropy}.
Nevertheless,  the closed string channel expansion is useful because we can easily evaluate the kinematic parts in this expression.
In fact, we only need the kinematic parts to study the quantities of interest (i.e., the area law term, the 
corner contribution, and the Markov gap).
In other words, for our purpose, the closed string channel expansion is more useful even though it involves the theory-dependent constants.

By this approximation, valid
in the limit $\beta/L\to 0$, 
each annulus amplitude can be approximated
as a product of
two strip amplitudes.
Correspondingly,
the total partition function
$Z_n$
can be approximated as a
product of two partition functions,
each obtained by gluing the strips,
\begin{align}
Z_n \equiv Z_{{\rm closed}}
\simeq \mathcal{N}^{-2}_n \cdot \mathcal{N}_n\cdot (Z_{{\rm open}})^2.
\end{align}
Here, 
the factor of $\mathcal{N}^{-2}_n$
comes 
from the normalization factor
while $\mathcal{N}_n$ comes from the coefficient $\ca{W}_{p_0}$.
Note that while we have the double insertions of the complete sets,
we do not need to square Eq.\ \eqref{eq:dof}
since $\braket{p_0|q_0} = \delta_{p_0,q_0}$. Each factor of $\mathcal{N}_n$ contributes to the entanglement entropy by $\lim_{n\ra 1}\frac{1}{1-n}\ln \mathcal{N}_n=-S_{\mathrm{topo}}$. 
Taking care of these contributions,
we recover the entanglement entropy
\eqref{top ent}:
\begin{equation}\label{eq:Sclosed}
\begin{aligned}
  S_{\mr{closed}}
  &=2 S_{\mr{open}}+2S_{\mr{topo}} - S_{\mr{topo}}\\
  &=2\pa{\frac{\pi cL}{48\beta}-S_{\mr{topo}}}+S_{\mr{topo}}\\
  &=\frac{\pi cL}{24\beta}-S_{\mr{topo}}.    
\end{aligned}
\end{equation}

Finally, we stress that using the open boundary condition introduces the cutoffs at the two boundaries
(at $-L/4$ and $L/4$) of the strip. After the conformal transformation $z\ra e^{z\frac{2\pi}{4\beta}}$, the cutoff is (see the right figure of Fig.\ \ref{fig:app}):
\begin{equation}
    \epsilon = e^{-\frac{L}{4}\frac{2\pi}{4\beta}}=e^{-\frac{\pi L}{8\beta}}.
\end{equation}
The area law term in $S_{\mr{open}}$ can be equivalently expressed in terms of the cutoff as:
\begin{equation}
    S_{\mr{open,area}}=\frac{\pi cL}{48\beta} =\fr{c}{6}\ln \epsilon^{-1}.
    \label{eq:s-area}
\end{equation}
This expression will be convenient in the following to take care of the
contribution to the entanglement entropy by conformal transformation. Namely,
the change of the entanglement entropy is encoded in the change of cutoff
$\epsilon\ra \epsilon'$. More generally, if the cutoffs on the left and right end of an interval are $\epsilon_1$ and $\epsilon_2$, respectively, the area law part can be expressed as:
\begin{equation}
    S_{\mr{open,area}} =\fr{c}{12}\ln \frac{1}{\epsilon_1 \epsilon_2}.
    \label{eq:s-area-diff}
\end{equation}
We also note that the topological term $S_{\mr{topo}}$
is independent of the cutoff $\epsilon$.

\section{Vertex States: 
Corner contributions to Entanglement Entropy}
\label{sec:Corner contributions and vertex states}

\begin{figure}[t]
  \begin{center}
    \includegraphics[width=7.0cm,clip]{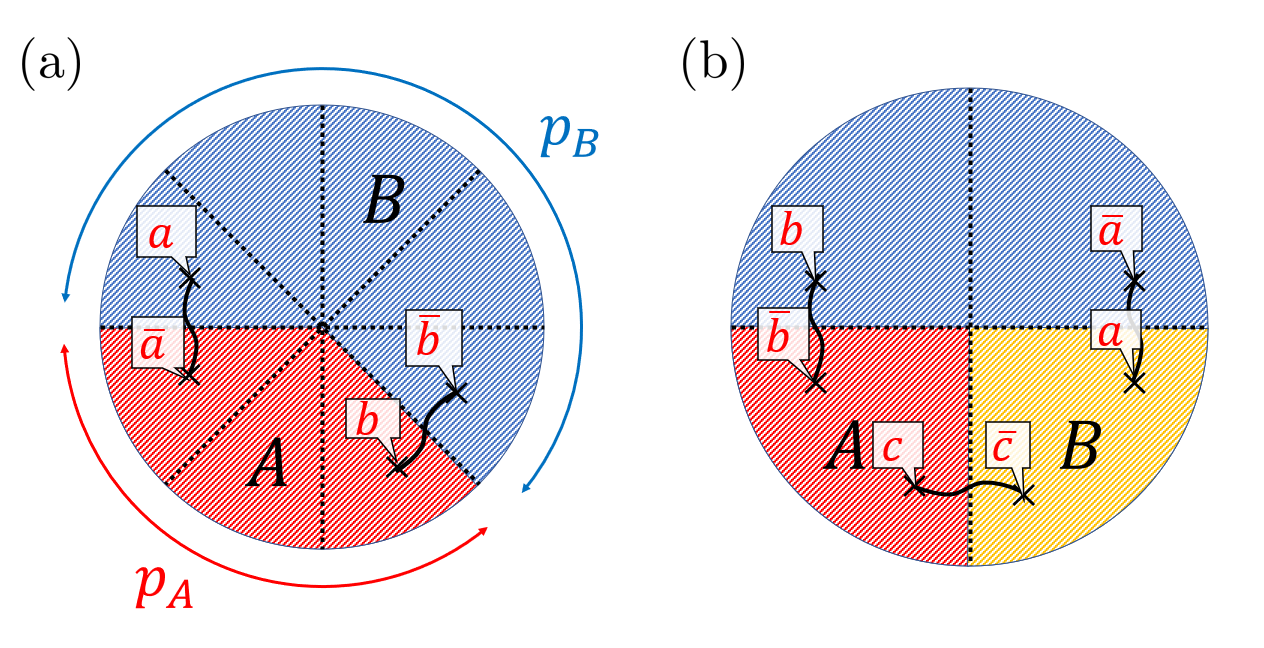}
  \end{center}
  \caption{
  (a) Bipartition setup:
  A spatial sphere divided into two regions $A,B$, and with four anyons ($a,\bar{a},b,\bar{b}$) insertion.
  We take the size of the subsystem as
  $p_A$ $(p_A + p_B = p)$,
  which leads to a corner with an angle
  $\theta = 2\pi p_A/p$.
  (b) Tripartition setup: A spatial sphere divided into three regions $A,B$ and $\overline{A\cup B}$, and 
  with six anyons ($a,\bar{a},b,\bar{b}, c, \bar{c}$) insertion.
  }
  \label{fig:anyon}
\end{figure}

We are now ready to generalize the previous discussions 
to more complicated setups. 
First, we can consider
the bipartition setup where we partition
the two-dimensional space into two regions $A$ and $B$
with a sharp corner (cusp)
[Fig.\ \ref{fig:anyon}(a)].
This type of bipartition was considered in
Refs.\ \cite{Rodr_guez_2010,
2021PhRvB.103k5115S, Ye_2022}
and a contribution to the entanglement entropy,
the so-called geometric or corner contribution,
has been identified. 
Second, we can consider
the multipartition setup where
we partition
the two-dimensional space into multiple regions $A, B, C,\ldots$
where all subregions meet at a junction (or junctions)
[Fig.\ \ref{fig:anyon}(b)].
This setup was discussed 
in Refs.\ \cite{2022PhRvB.105k5107L, siva2022universal},
and the reflected entropy (the Markov gap) and entanglement negativity
were computed. 
We note that the first setup can be obtained
from the multipartition setup by simply
grouping 
the multiple regions $A, B, C,\ldots$
into two groups and regarding the regions in the same group
belonging to the same Hilbert space.

In this section, we focus on 
the 
corner contribution 
to bipartite entanglement entropy.
To provide a unified treatment, 
we will consider the setup in 
Fig.\ \ref{fig:anyon}(b)
throughout the paper,
in which we $p$-partition 
the spatial sphere,
and consider two regions $A$ and $B$
(and the rest).
The corner angle of the region $A$
is denoted by $\theta = 2\pi p_A/p$
($1< p_A \le p$).
We limit out discussion to the case where $A$ and $B$ are adjacent, unless specified otherwise. 
We also insert pairs of anyons
($a$ and $\bar{a}$, etc.)
across the interfaces (entangling boundaries)
that give rise to extra topological contributions to bipartite entanglement entropy,
while they have nothing to do with the geometrical contribution.



\subsection{Vertex states}
\label{sec:def-vertex-state}

As noted in Ref.\ \cite{2022PhRvB.105k5107L},
using the bulk-boundary correspondence, 
the multipartition setup can be related to
vertex states in CFT.
Specifically, the topological ground states near the
multipartite entangling boundary can be approximated by
a vertex state.
For a given (chiral) CFT, a $p$-vertex state $|V\rangle$
can be defined as follows.
Here, we focus on chiral topological order and hence chiral
CFT (chiral edge states).
Analogously to boundary states,
a vertex state $|V\rangle$ defined in the tensor product
of $p$-copies of the CFT (defined on a spatial circle of length $L$)
satisfies 
\begin{equation}
  \label{vertex state cond 1}
  [T^i(\sigma)-T^{i+1}(L-\sigma)]|V\rangle=0,\quad 0\leq \sigma\leq L/2,
\end{equation}
where $T^{i}(\sigma)$ is the stress-energy tensor
of the $i$-th copy ($i=1,\ldots, p$),
and $\sigma$ coordinatizes the spatial circle. 
If there is a (conserved) current in CFT,
vertex states satisfy,
additionally, 
a condition similar to 
\eqref{vertex state cond 1}
in terms of the current.
For example, for the free Majorana fermion CFT,
a vertex state satisfies 
\begin{equation}
  [\psi^{i}(\sigma)+i\psi^{i+1}(L-\sigma)]|V\rangle=0,\quad 0\leq \sigma\leq L/2.
  \label{eqn:usual}
\end{equation}
where $\psi^i$ is the $i$-th copy of the Majorana fermion field.

Vertex states can be constructed explicitly for
non-interacting theories, the real or complex free fermion theory,
by solving Eq.\ \eqref{eqn:usual} explicitly.
It is then possible to compute the
various entanglement measures directly,
following the spirit of 
Sec.\ \ref{sec:Boundary states and left-right entanglement entropy}.
This calculation was carried out in
Ref.\ \cite{2022PhRvB.105k5107L}.
We will also revisit this calculation in
Sec.\ \ref{sec:numerics}
for the case of four vertex states ($p=4$).
For general CFTs (RCFTs),
on the other hand, 
finding vertex states 
is a yet challenging task.
However,
as we will show below, 
by using the path integral representation
and conformal interface approach,
we can still obtain the universal behaviors
of its entanglement measures.

\begin{figure}[t]
  \begin{center}
    \includegraphics[width=\linewidth]{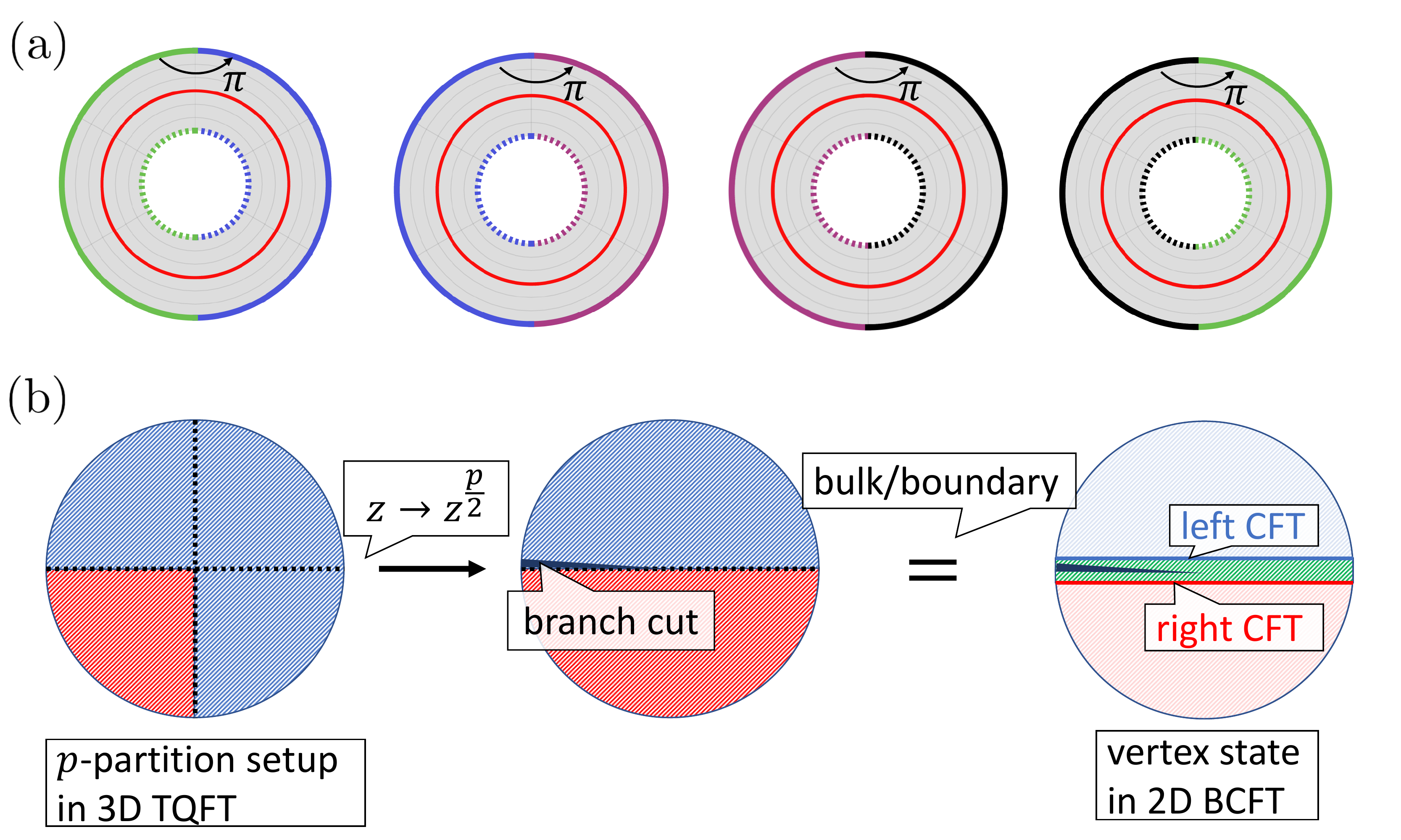}
  \end{center}
  \caption{(a) Illustration of the excess angle of 
  the $p$-vertex state density matrix path integral. As shown in the figure, on the outer edge of the top of the annulus, the total angle is $p\pi$, thus the excess angle is $p\pi-2\pi=2\pi(\frac{p}{2}-1)$. Similarly, the excess angles on the inner edge of the top, and the outer and inner edge of the bottom of the annulus are also $2\pi(\frac{p}{2}-1)$.  (b) The $p$-partition setup of the two-dimensional spatial sphere, where the multipartite entanglement measures are computed. 
  To obtain the corresponding $p$-vertex state on the entanglement boundary, we first map the spatial sphere (Left) into the Riemann surface with exceed angle $2 \pi \pa{\fr{p}{2}-1}$ (Center) by conformal transformation $z\ra z^{p/2}$.
    Applying the bulk-boundary correspondence to this surface, we obtain the vertex state (Right), whose density matrix is defined on spacetime manifold with four exceed angles, each taking the value $2\pi(\frac{p}{2}-1)$. 
    }
  \label{fig:vertex}
\end{figure}

We have to mention that the $p$-vertex state is defined \textit{after} the conformal map (see Fig. \ref{fig:vertex}),
\begin{equation}\label{eq:p}
z \to z^{\fr{p}{2}}.
\end{equation}
The reason is as follows. On the one hand, the multipartite entanglement measures are computed on the two-dimensional spatial sphere with a $p$-partition setup. On the other hand, by definition, the density matrix of $p$-vertex state is defined on the spacetime manifold with four excess angles, each takes the value $2\pi(\frac{p}{2}-1)$ as shown in Fig.\ \ref{fig:vertex} (a) (and after taking the trace of $\rho$, the spacetime path integral becomes a $p-1$ genus). Thus, the conformal transformation in Eq.\ \eqref{eq:p} is required to bridge them, as shown in Fig.\ \ref{fig:vertex} (b). 
If this is not included, the final results for all entanglement measures would carry an extra factor of $\frac{c}{3}\ln \frac{p}{2}$.

\label{sec:cft-vertex}

\subsection{Entanglement entropy with corner contribution}
\label{sec:vertex-entanglement}

\begin{figure}[t]
  \begin{center}
    \includegraphics[width=4.0cm,clip]{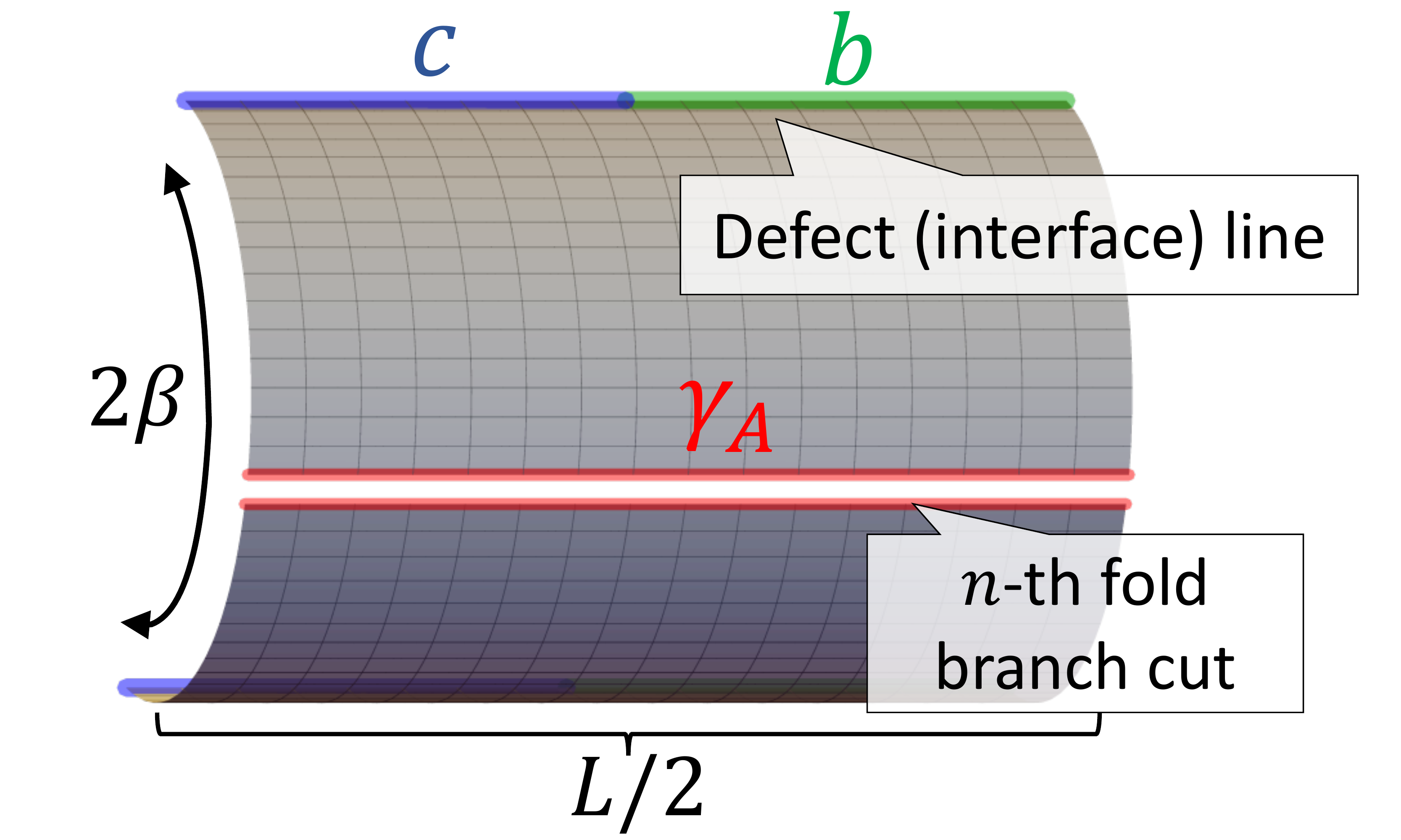}
    \includegraphics[width=4.0cm,clip]{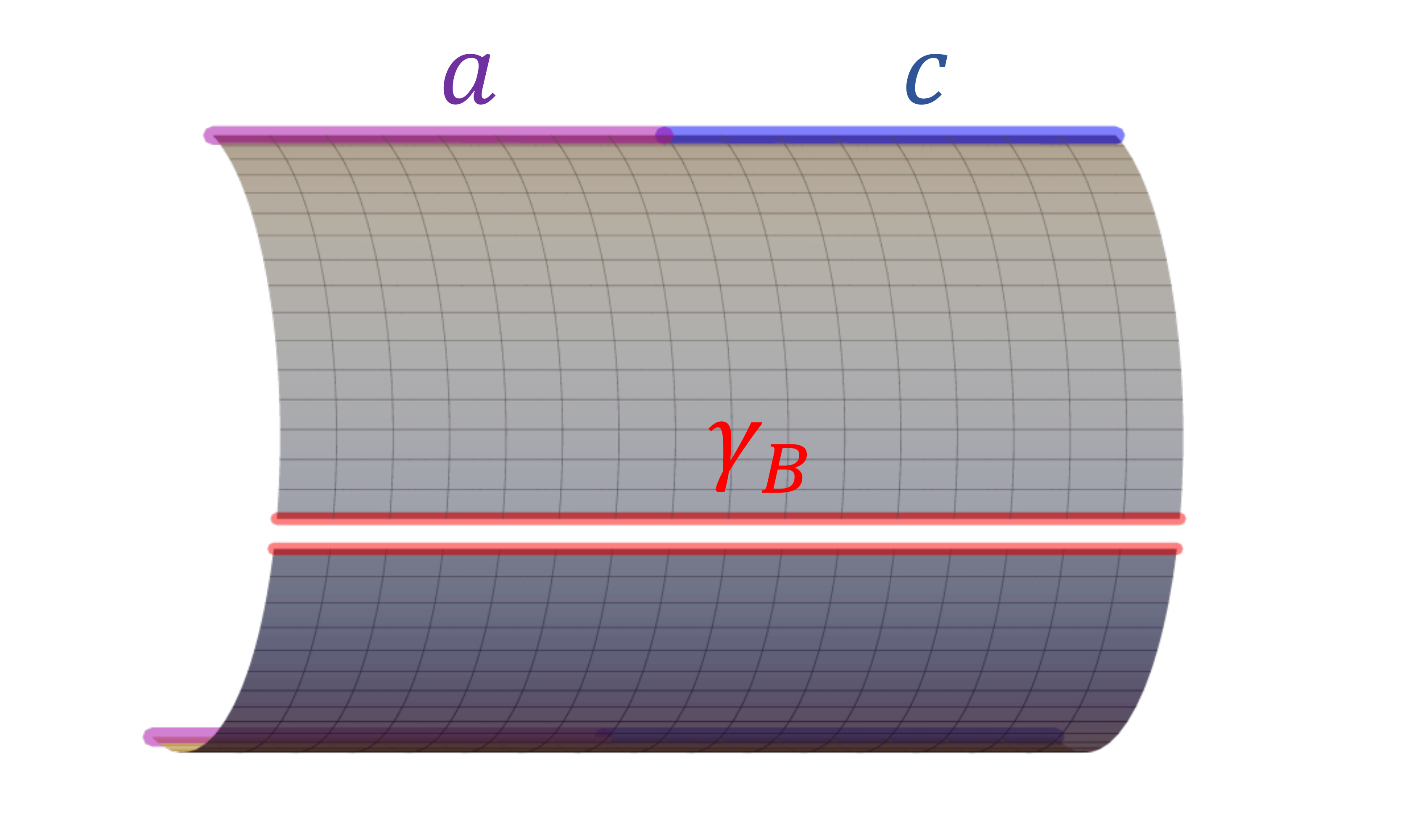}
    \includegraphics[width=4.0cm,clip]{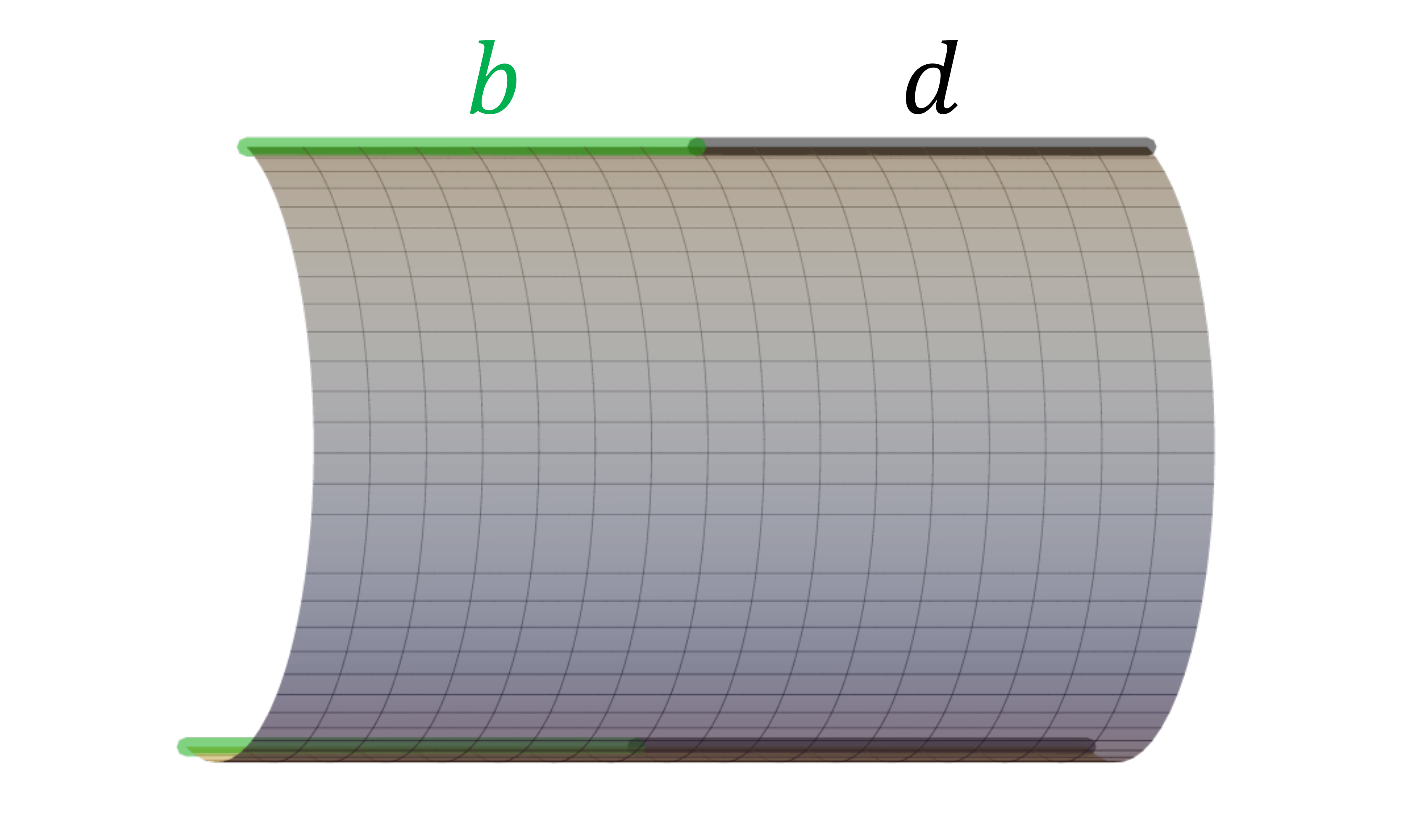}
    \includegraphics[width=4.0cm,clip]{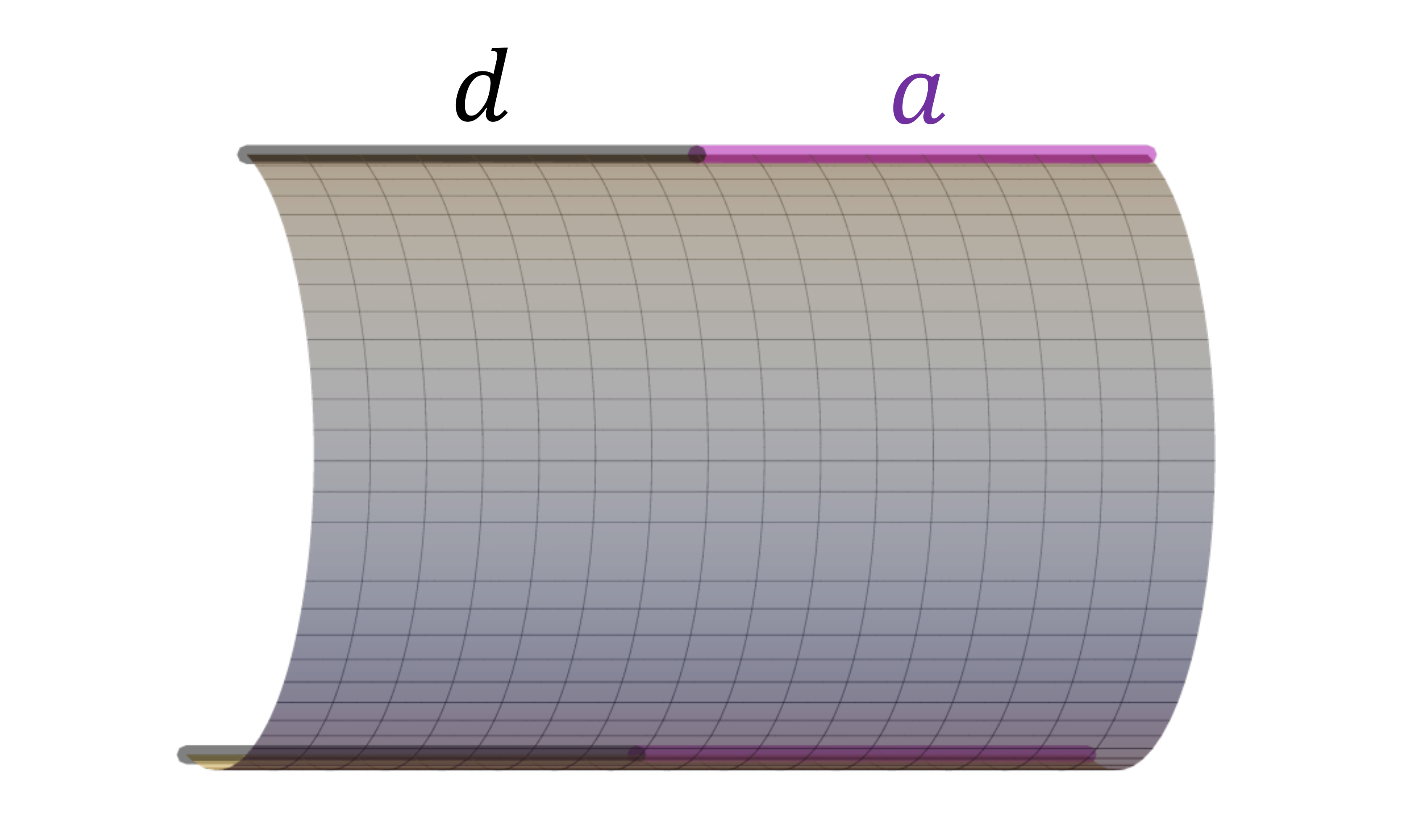}
  \end{center}
  \caption{Sketch of our setup for $4$-vertex state partition function with open boundary condition. 
  The width of each strip is $2\beta$ and the length is $L/2$. 
    When evaluating $n$-th R\'enyi entropy, the intervals $\gamma_A$ and $\gamma_B$ denote the $n$-th fold branch cut.
    On the top and bottom, we have the excess angle $2\pi(\frac{4}{2}-1)=2\pi$.
    }
  \label{fig:setup}
\end{figure}

We are now ready to calculate the entanglement entropies
$S_A$, $S_B$, and $S_{AB}$ associated to
the regions $A$, $B$, and $AB$.
in Fig.\ \ref{fig:anyon}(b).
The multipartition can be done by
using the $p$-vertex state.
The corresponding replica path integral for computing $n$-th R\'enyi entropy
is defined on a Riemann surface.
For example, when $p_A=1$, the path integral is on a Riemann surface with genus $n(p-2)+1$. The path integral on the higher genus surface is not readily obtained. 
However, 
in the limit of $\beta/L\to 0$, it
can be evaluated by using 
the path integral decomposition
described in Sec.\ \ref{sec:path-int-dec}.
Here,
we factorize the partition function (with closed boundary condition)
into two 
with open boundary condition by inserting a complete set of states.
In the limit $\beta/L \to 0$,
the open partition function can
be approximated by
taking the leading twisted operator or vacuum.
Thus, the replica partition function $Z_{\text{closed}}$ can be approximated by
$Z_{\text{closed}}\approx \mathcal{N}^{-1} \pa{Z_{\text{open}}}^2$.
The constant $\mathcal{N}$ is a combination of the normalization factor and the
coefficient $\ca{W}_p$, which contributes to the entanglement entropy as the
topological entropy, as discussed in Eq.\ \eqref{eq:Sclosed}.
This strip partition function can be illustrated
as in Fig.\ \ref{fig:setup} for the case of $p=4$, where the width and length of each strip are $2\beta$ and $L/2$, respectively.
We take two intervals $\gamma_A$ and 
$\gamma_B$ as shown in the figure, 
namely, the two bulk regions $A,B$ are adjacent.  The intervals $\gamma_A$ and $\gamma_B$ denote the $n$-th fold branch cut when evaluating $n$-th R\'enyi entropy.

The entanglement entropy $S_A$ and $S_{AB}$ can be computed by the open partition function $Z_{\mathrm{open}}$. To evaluate the partition function $Z_{\mathrm{open}}$, we consider the following conformal maps (the following figures are illustrated for $p=4,p_A=p_B=1$):

\begin{enumerate}[(i)]
  \setlength{\itemsep}{0cm}

\item Mapping from cylinder to hemisphere: $z \to \ex{-\fr{\pi}{2\beta} z}$. 
\label{eq:i}

\item Rotating  : $z \to \fr{1+z}{1-z}$.
\label{eq:ii}

 (insertion points of twist operators $(0,\infty)$ mapped to $(1,-1)$ )

\item Unwrapping: $z \to z^{\fr{2}{p}}$ and gluing $p$ copies.
\label{eq:iii}
\begin{figure}[H]
\centering
  \includegraphics[width=5.0cm,clip]{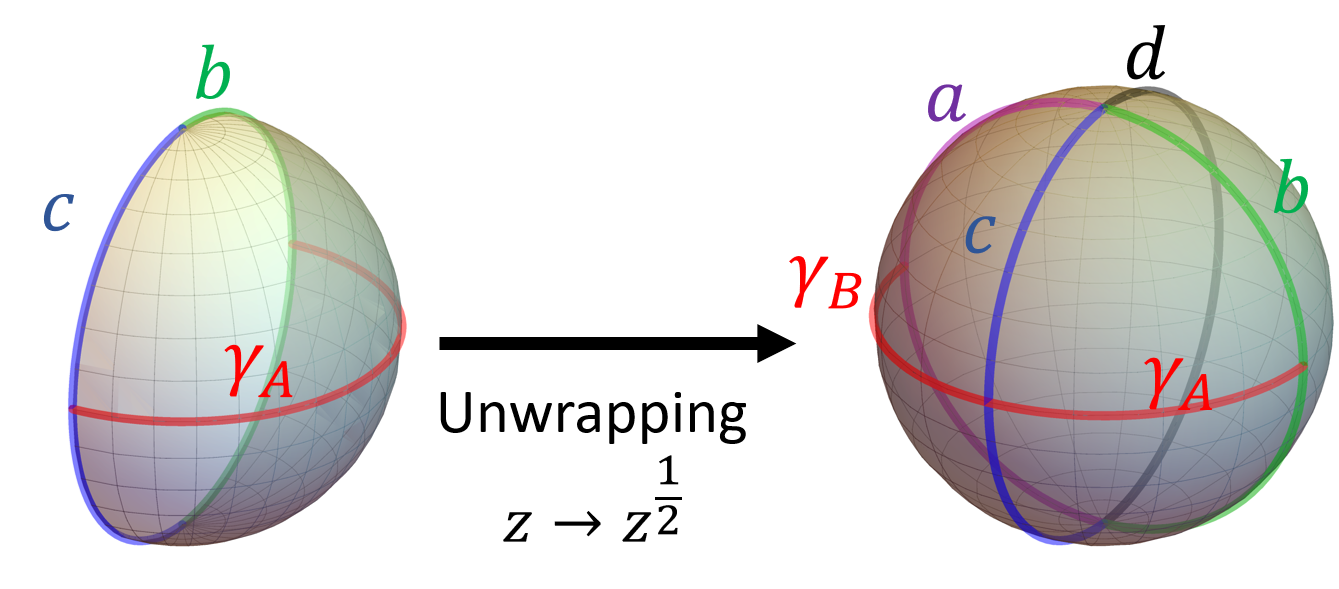}
\end{figure}
\vspace{-0.3cm}

\item Rotating    : $z \to -\ex{\fr{2 \pi i}{p}} \fr{z - \ex{-\fr{2 \pi i}{p}}  }{  z - \ex{\fr{2 \pi i}{p}}  }   $.
\label{eq:iv}

 (insertion points of twist operators $(\ex{-\fr{2 \pi i}{p}} ,1, \ex{\fr{2 \pi i}{p}} )$ mapped to $(0,1,\infty)$ )
 
 \item Mapping from sphere to cylinder $z\ra \ln z$.
 \begin{figure}[H]
     \centering
     \includegraphics[width=0.45\linewidth]{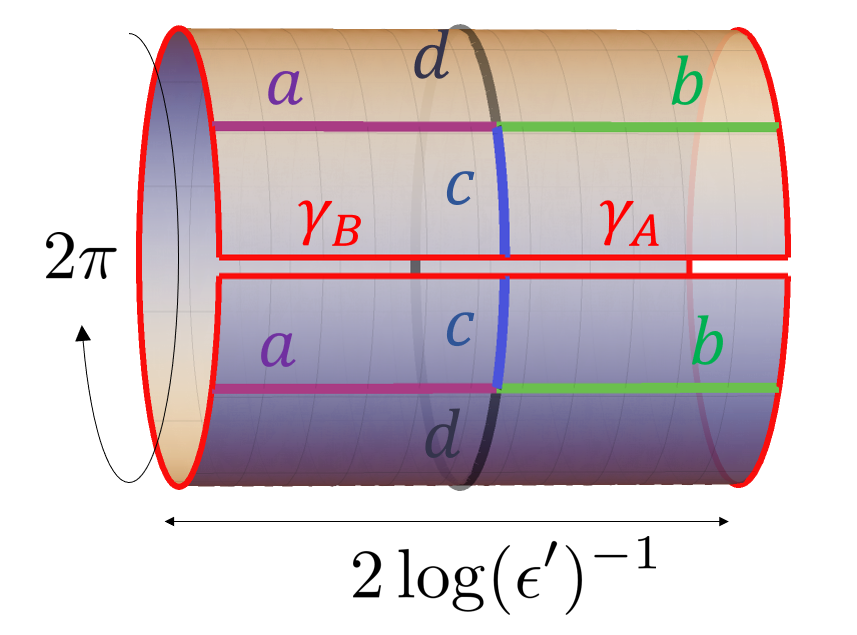}
 \end{figure}
 \vspace{-0.5cm}
 \end{enumerate}

 In Step \eqref{eq:iv},
 the interval $\gamma_B$ is mapped to $(0,1)$ and the interval $\gamma_A$ is mapped to $(1,\infty)$. 
 The cutoff
 $\epsilon=e^{-\frac{\pi L}{8\beta}}$ 
 at
 the end of the intervals ($z=0,1,\infty$)
 transforms as
\begin{equation}
\begin{aligned}
   \epsilon \to \epsilon'=\fr{2}{p\sin\fr{2\pi }{p}} \epsilon     ,&  \ \ \  \text{at }  z=0  ,\\
   \epsilon \to \epsilon'=\frac{4}{p \tan \frac{\pi}{p}}\epsilon, & \ \ \ 
   \text{at } z= 1,
   \\
   \fr{1}{\epsilon} \to \fr{1}{\epsilon'}=\fr{p\sin\fr{2\pi }{p}}{2} \fr{1}{\epsilon}     ,&    \ \ \ \text{at } z=\infty. \\
\end{aligned}
\label{eq:epsilon1-4}
\end{equation}
The interval $\gamma_A\cup \gamma_B$ 
is mapped to $(0,\infty)$. By using Eq.\ \eqref{eq:s-area} with the modified
cutoffs in Eq.\ \eqref{eq:epsilon1-4}
($S_{\mr{open,area}}=\frac{c}{6}\ln(\epsilon')^{-1}$), and the normalization
factor from Eq.\ \eqref{eq:norm-s} and the coefficient $\ca{W}_{p_0}$ from
Eq.\ \eqref{eq:dof}, the entanglement entropy $S_{AB}$ including the corner contribution is given by
\begin{equation}
\begin{aligned}
  S_{AB}
  &=2S_{\mr{open}}+S_{\mr{topo}}
  \\
  &= 2\left(
    \frac{\pi c L}{48\beta}+\frac{c}{6}\ln \sin\fr{2\pi}{p}
    \right)-S_{\mr{topo}}
  \\
    & = \frac{\pi c L}{24\beta}+\frac{c}{3}\ln \sin\fr{2\pi}{p}+\ln \frac{d_a d_b}{\cal{D}}.
\end{aligned}
\end{equation}
The topological term comes from the anyons insertion $a,\bar{a}$ and $b,\bar{b}$, 
where unique fusion channel is assumed
(as commented in the last paragraph of Sec.\ \ref{sec:Boundary states and left-right entanglement entropy}).
We set $d$ as the trivial interface from now on. 
In this expression the factor $\frac{c}{3}\ln \fr{p}{2}$ is removed
by Eq.\ \eqref{eq:p}. 
Similarly, to find the entanglement entropy for interval $\gamma_A$, we can perform an
additional shift $z\ra z-1$ such that
the interval $\gamma_A$ is mapped to $(0,\infty)$ and apply Eq.\ \eqref{eq:s-area-diff}:
\begin{equation}
\begin{aligned}
  S_A&=
       \frac{\pi c L}{24 \beta}+\frac{c}{6}\ln
       \left[\frac{1}{2}\sin\frac{2\pi}{p}\tan \frac{\pi}{p}
       \right]+\ln \frac{d_b d_c}{\mathcal{D}}\\
     & =
       \frac{\pi c L}{24 \beta}+\frac{c}{3}\ln \sin\frac{\pi}{p}+\ln \frac{d_b d_c}{\mathcal{D}}.
\end{aligned}
\end{equation}
And similarly, for $S_B$,
\begin{equation}
\begin{aligned} 
  S_B &=
        \fr{\pi c L}{ 24\beta}   + \fr{c}{3} \ln \sin\fr{\pi }{p}  +  \ln \fr{d_a d_c}{\ca{D}}.
\end{aligned}
\end{equation}
Combining these results,
we can also obtain the mutual information:
\begin{equation}\label{eqn:mi-p-vertex}
\begin{aligned}
I(A,B)
&= S_A + S_B -S_{AB}    \\
&=
  \fr{ \pi c L}{24 \beta}   +   \fr{c}{3} \ln \fr{ \tan\fr{\pi}{p}}{2}+\ln \frac{d_c^2}{\cal{D}}.
\end{aligned}
\end{equation}

The above results, presented for $p_A=1$,
can be readily generalized to $1< p_A \le p$.
I.e., subregion $A$ has a cusp with
angle
$\theta = 2\pi p_A/p$.
The entanglement entropy $S_A$ in this case is given by
\begin{equation}
  S_A = \frac{\pi c L}{24\beta}
  +\frac{c}{3}\ln \sin \frac{\theta}{2}
  -S_{\mr{topo}}.
  \label{eqn:ent-p-vertex}
\end{equation}
This is the central result of this section. 
The second term
$
\frac{c}{3}\ln \sin \frac{\theta}{2}
$
can be identified as
the corner contribution 
to the entanglement entropy. 
Recalling that we have two 
corners with equal angles
in our setup,
we may introduce  
$
a(\theta):=
-(1/2) \times \frac{c}{3}\ln \sin \frac{\theta}{2}
$
that represents contribution 
from each corner 
(with the minus sign to
be consistent with the convention 
in 
\cite{Rodr_guez_2010,2021PhRvB.103k5115S, Ye_2022}.
We note that 
$-2a(\theta)$
takes the same form as
the entanglement entropy of the ground state
of (1+1)d CFT on a finite periodic chain (of length $p$)
associated
with an interval (of length $p_A$).
It is known that 
the same contribution arises 
when the subregion
of our interest includes 
a physical edge 
\cite{Estienne_2020},
although there is no physical 
edge in our setup. 

In Sec.\ \ref{sec:numerics}, 
we will compute the corner contribution
for the case of the free Majorana fermion CFT ($c=1/2$)
and $p=4$.
There, we will construct the $p=4$ vertex states explicitly,
and calculate the bipartite entanglement entropy
numerically. 
We will confirm, within numerical errors,
the above result \eqref{eqn:ent-p-vertex}.

We also note that the corner  
contribution $a(\theta)$
was previously discussed in the literature for 
the integer quantum Hall ground
states
\cite{Rodr_guez_2010,2021PhRvB.103k5115S, Ye_2022}.
In \cite{2021PhRvB.103k5115S},
it was numerically observed that
the corner contribution (from one corner)
behaves as $a(\theta) \sim \kappa/\theta$
for small $\theta$
while $a(\theta) \sim \sigma (\theta-\pi)^2$
for $\theta \sim \pi$,
where numerical constants 
$\kappa$ and $\sigma$ were determined numerically.
While the latter behavior 
of the corner contribution 
near $\theta\sim \pi$
is consistent with ours,   
the small $\theta$ behavior 
disagrees 
with $(-c/6)\ln \sin \theta/2 \sim (-c/6) \ln \theta$. 
We also note that 
in \cite{2021PhRvB.103k5115S},
the constant 
$\sigma$ is estimated 
for the $\nu=1$ integer quantum Hall state
as $\sigma\sim 0.02836$,
that should be contrasted with
$
(-c/6) \ln \sin \theta/2 
\sim 
(c/48) (\theta-\pi)^2
$
with $c/48\sim 0.02083$ for $c=1$. 
The source of the discrepancy is not entirely clear. 
We nevertheless recall that 
the (single-particle) 
bipartite entanglement spectrum 
for the integer quantum Hall state
in the lowest Landau level
is linear only for small momentum 
along the entangling boundary
\cite{Rodr_guez_2009},
while in our ansatz state 
\eqref{initial density matrix}
a perfectly relativistic spectrum
is assumed.
It is also possible that for small enough $\theta$,
the representation of the 
topological ground state near the cusp
by using the vertex state,
$e^{- \beta H_0}|V\rangle$,
is not entirely accurate.
Our numerics in Sec.\ \ref{sec:numerics}, where we verify the formula 
\eqref{eqn:ent-p-vertex}
for relatively large $\theta$
by using the explicit form of 
$e^{- \beta H_0}|V\rangle$,
is in favor of these speculations.
Finally,
we should also note that
not all
bulk geometrical 
properties of topological liquid
may be captured 
by using the edge states
and the bulk-boundary correspondence
\cite{Gromov_2016}.

\section{Reflected entropy and Markov gap}

\label{sec:reflected-entropy}


In this section, we consider the multipartition setup as in Fig.\ \ref{fig:anyon}(b) and compute the reflected entropy as well as the Markov gap for the $p$-vertex state for the case where $A$ and $B$ are adjacent and $p_A=p_B=1$. We will show that the Markov gap is $\frac{c}{3}\ln 2$ analytically. 

Let us first define the reflected entropy and its replica path integral. 
The reflected entropy is a correlation measure that captures the tripartite
entanglement. Given a reduced density matrix
$\rho_{A\cup B}$ supported on $A\cup B$,
one can obtain its canonical purification state
$|\sqrt{\rho}\rangle\!\rangle$
which is supported on $A \cup B \cup A^* \cup B^*$,
and
$A^*$, $B^*$ are identical copies of $A, B$
(up to complex conjugation).
The reflected entropy $S_R$ is defined as the von Neumann entanglement entropy
of the state $|\sqrt{\rho}\rangle\!\rangle$
after tracing out $B\cup B^*$:
\begin{align}
  S_R = S(\rho_{A\cup A^*}),
  \quad
  \rho_{A\cup A^*} =
  \mathrm{Tr}_{B\cup B^*}\, 
  |\sqrt{\rho}\rangle\!\rangle
  \langle \!\langle \sqrt{\rho}|
\end{align}
In the following, we will use the replica trick to compute the reflected entropy. 

To compute the reflected entropy, two replica indices $m,n$ shall be introduced and the open partition function to be evaluated is:
\begin{equation}
Z_{n,m}=\tr_{AA^*} \left( \tr_{BB^*}|\rho_{AB}^{m/2}\rangle\!\rangle\langle\!\langle\rho_{AB}^{m/2}|\right)^n.
\label{eqn:Znm}
\end{equation}
Here $n$ is the index for R\'enyi replicas, and $m$ is the index for handling the square root of the reduced density matrix, where we take $m\ra 1$ at the end of the calculation.
The $n$-th R\'enyi reflected entropy is computed using:
\begin{equation}
    S_R^n = \lim_{m\ra 1}\frac{1}{1-n}\ln \frac{Z_{n,m}}{(Z_{1,m})^n},
\end{equation}
and the reflected entropy is $S_R=\lim_{n\ra 1}S_R^n$.

Similar to the previous discussion, the path integral for evaluating $S_R^n$ is performed on a 
higher genus surface and is hard to compute. Using the 
path-integral decomposition as in Sec.\ \ref{sec:path-int-dec}, we again focus on the open partition function.
For reflected entropy, the normalization factor and the density of the lowest energy states become highly complicated.
Therefore,  we first focus on the special case 
where the topological term can be neglected and give a few comments later.

To evaluate the partition function $Z_{n,m}$,
we consider the following conformal maps. The steps (i)-(iv) are the same as in Sec.\ \ref{sec:vertex-entanglement}, and the conformal maps from step (v) are:

\begin{enumerate}[(i)]
\setcounter{enumi}{4}
  \setlength{\itemsep}{0cm}

\item Unwrapping: $z \to z^{\fr{1}{m}}$.
\label{eq:v}
\begin{figure}[H]
\centering
\includegraphics[width=5.0cm,clip]{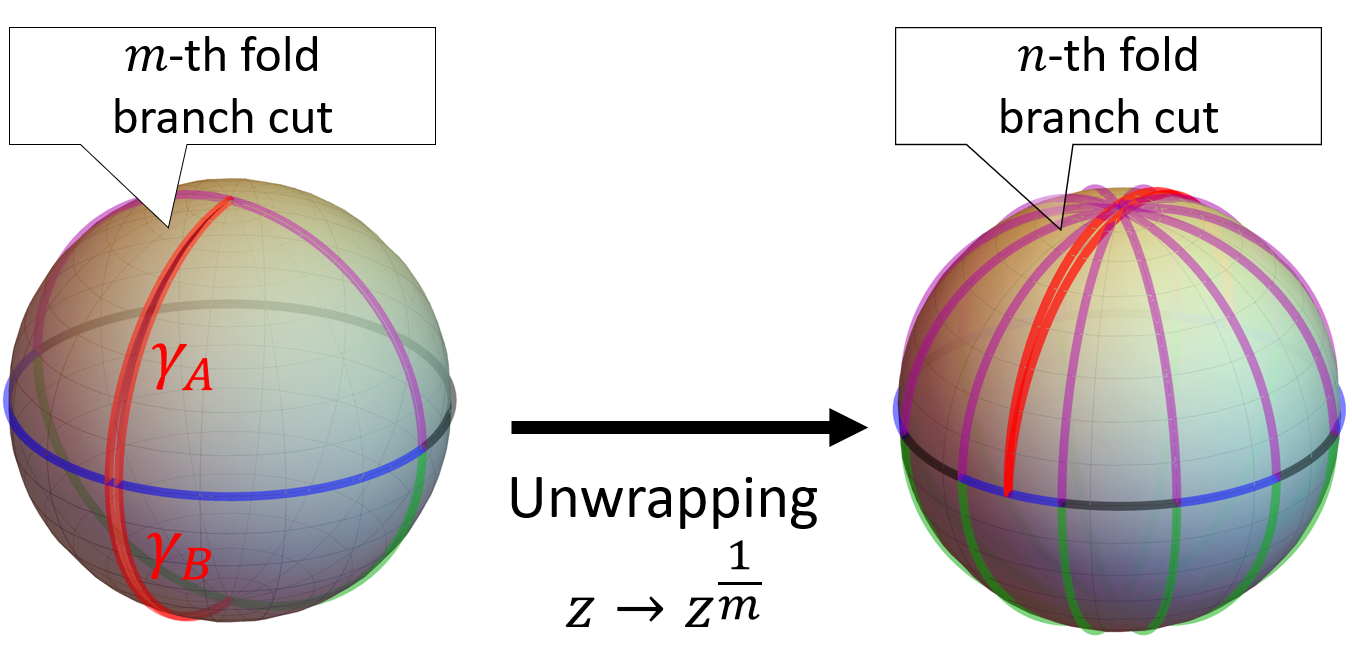}
\end{figure}

\item Rotating  : $z \to \fr{1+z}{1-z}$.
\label{eq:vi}

 (insertion points of twist operators $(-1,1)$ mapped to  $(0,\infty)$ )

\item Unwrapping: $z \to z^{\fr{1}{n}}$.
\label{eq:vii}

\item Map from sphere to cylinder: $z \to \ln z$.
\begin{figure}[H]
\centering
\includegraphics[width=5.0cm,clip]{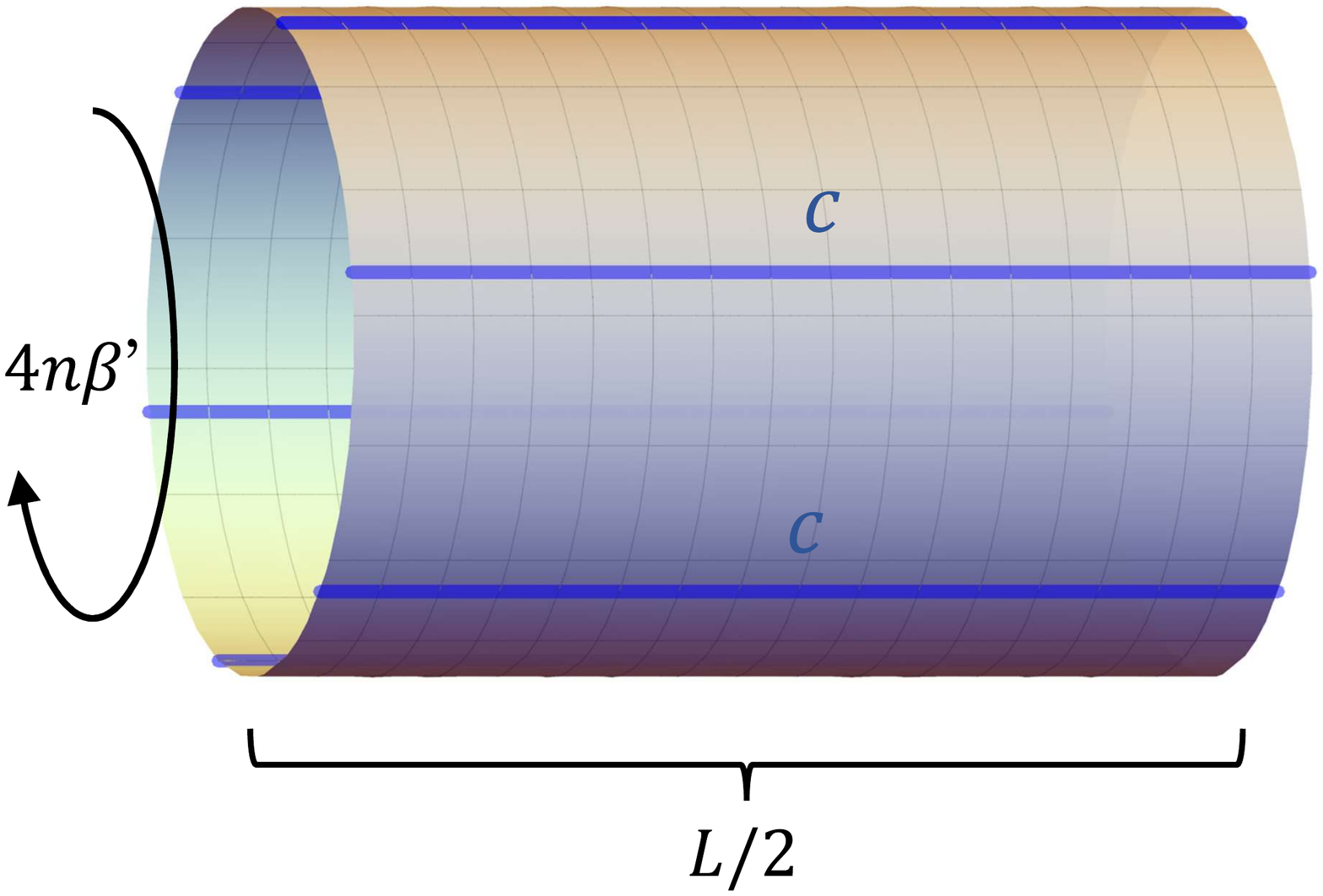}
\end{figure}
\vspace{-1cm}
\label{eq:viii}

\end{enumerate}

\noindent



After the above conformal transformations, $Z_{n,m}$ is brought to a partition function on a cylinder with interfaces, and with a new cutoff:
\begin{equation}
    \epsilon \ra  \epsilon'=\frac{1}{2m}\fr{4}{p \tan \frac{\pi}{p}}\epsilon.
    \label{eq:epsilon-prime}
\end{equation}
This conformal factor $\pa{2m}^{-2h_n}$ is the OPE coefficient $C_{\sigma{g_A} \sigma{g^{-1}_B} \sigma{g^{-1}_A g_B} }$ of twist operators \cite{2021JHEP...03..178D}.
Since we assume the interfaces to be topological,
we can freely move junction fields so that the net of the interfaces is simplified as shown in step (\ref{eq:viii}), where $\beta'=-\frac{\pi L}{8\ln(\epsilon')}=\beta (1+\fr{8\beta}{\pi L}\ln \fr{1}{2} \fr{4}{p \tan\fr{\pi}{p}} + O(\beta^2))$.


Using the new cutoff $\epsilon'$, the reflected entropy can be evaluated by Eq.\ \eqref{eq:s-area}:
\begin{equation}
\begin{aligned}
    S_R&=\frac{c}{3}\ln {\epsilon'}^{-1}\\
    & =\lim_{m\ra 1}\fr{c}{3}\left[
    \ln \epsilon^{-1}+\ln 2m +\ln \frac{p \tan\frac{\pi}{p}}{4}
    \right],
\end{aligned}
\end{equation}
where we use the assumption that the topological contribution is zero.
Taking into account the effect of the conformal map (\ref{eq:p}),
we obtain the reflected entropy for the $p$-partition setup,
\begin{equation}
  \label{eq:SR}
S_R(A,B) = \fr{c}{24} \fr{\pi L}{\beta}   +   \fr{c}{3} \ln \fr{\tan\fr{\pi}{p}}{2}  + \fr{c}{3}\ln 2.
\end{equation}
The first term corresponds to the area law term.
The second term comes from the conformal transformation (\ref{eq:epsilon1-4}).
This term can be understood as 
a corner contribution. 
The third term comes from the $(2m)^{-1}$ factor in (\ref{eq:epsilon-prime}).
This is the universal contribution related to the OPE coefficient $C_{\sigma{g_A} \sigma{g^{-1}_B} \sigma{g^{-1}_A g_B} }$.



With the above results on mutual information and reflected entropy, let us now evaluate the Markov gap. Combining the results from the previous two subsections and assuming the topological contribution can be neglected, we find the Markov gap is:  
\begin{equation}
  \label{Markov gap result}
h=S_R(A,B) - I(A,B) = \fr{c}{3} \ln 2.
\end{equation}
In summary, the corner 
contribution from mutual information and reflected entropy cancel out exactly, and the Markov gap does not receive 
the corner contribution. 
This is verified numerically in the Majorana fermion CFT in Sec.\ \ref{sec:numerics}. 


Now, to reiterate, we restricted our computation to the case in which the reflected entropy does not receive additional topological contributions from anyon insertions. In previous sections, we considered configurations in which a single pair of anyons pierces each interface, with the anyon label being determined by the corresponding interface operator, as depicted in Fig. \ref{fig:anyon}(b). While we have been unable to compute the Markov gap in this more general case, we suspect that it should still vanish. Indeed, as noted in the Introduction, Ref.\ \cite{2021PhRvL.126l0501Z} proved that the Markov gap vanishes for so-called ``sum-of-triangle" states. For a tripartition of a Hilbert space $\mathcal{H} = \mathcal{H}_A \otimes \mathcal{H}_B \otimes \mathcal{H}_C$ and further bipartitions of each subspace, $\mathcal{H}_\alpha = \bigoplus_j \mathcal{H}_{\alpha_L^j} \otimes \mathcal{H}_{\alpha_R^j}$ a sum of triangle states takes the form
\begin{align*}
	\ket{\psi} = \sum_j \sqrt{p_j} \ket{\psi_j}_{A_R^j B_L^j} \ket{\psi_j}_{B_R^j C_L^j} \ket{\psi_j}_{C_R^j A_L^j}
\end{align*}
where $\sum_j p_j = 1$ and $\ket{\psi_j}_{\alpha_R^j \beta_L^j}$ has support in $\mathcal{H}_{\alpha_R^j} \otimes \mathcal{H}_{\beta_L^j}$. Qualitatively, the anyon configuration Fig.\ \ref{fig:anyon}(b) takes this triangle state form, with each anyon pair entangling two of the three subregions.
Based on this heuristic, we suspect that this particular configuration of anyons will not lead to a non-zero Markov gap. It is possible that other, non-trivial anyon configurations could lead to a non-vanishing Markov gap.


\section{Four-vertex states in the Majorana fermion
  CFT and correlation measures}
\label{sec:numerics}

In this section,
we consider the free real fermion CFT with $c=1/2$,
and construct four-vertex states explicitly. 
From
the explicit form of the vertex states, 
various correlations measures
(entanglement entropy, reflected entropy,
entanglement negativity, etc.)
can be calculated numerically. 
We will see that the numerics
is consistent with
the analytical results in the preceding sections. 
In addition, we can calculate
the correlation measures in
the setups that are not amendable in the analytical treatment.
For example,
we will discuss the correlation measures
when subregions 
$A$ and $B$ are not adjacent.

%

In the following, we will present two
approaches to construct vertex states;
the direct method and the Neumann function method \cite{1987NuPhB.293...29G,1989NuPhB.317..411L,2006PThPh.115..979I,2008PThPh.119..643I,2022PhRvB.105k5107L}. 
We will also note that there are at least two vertex states,
which satisfy what we call ``usual'' and ``kink'' boundary conditions.
We will show that
the Neumann function method,
when applied naively, 
gives rise to the vertex state with the kink boundary condition,
and the kink boundary condition reproduces the correlation measures as predicted
in the previous sections.


\subsection{Usual and kink boundary conditions}

Let us consider four copies of the Majorana fermion CFT and construct the vertex state. 
We denote the Majorana fermion fields
by $\psi^i(\sigma)$ where $i=1,\ldots,4$ denotes the copy index, 
and $0 \le \sigma \le 2\pi$
coordinatizes the spatial circle.
For simplicity in this section we set the circumference of the circle
to be $L=2\pi$.
They satisfy the canonical anticommutation relation
$\{\psi^i(\sigma), \psi^j(\sigma')\}=
2\pi \delta^{ij}
 \sum_{n \in \mathbb{Z}} \delta(\sigma-\sigma'-2\pi n)$. 
We will work with the Neveu-Schwartz (antiperiodic) boundary condition in $\sigma$.
Under the antiperiodic boundary condition, the fermion field can be expended in the Fourier modes labeled by half-integers as
$
\psi^i(\sigma)
=
\sum_{r\in \mathbb{Z}+1/2} \psi^i_{r}e^{ -i r \sigma}
$. Using the Fourier modes, the canonical anticommutation relation reads $\lbrace \psi^i_r,\psi^j_{s}\rbrace=\delta^{ij}\delta_{r+s,0}$. 

As discussed in Sec.\ \ref{sec:def-vertex-state},  
we define a vertex state $|V\rangle$
in terms of the boundary condition
it satisfies. First,
we introduce the usual boundary condition by
\begin{align}
  \label{usual cond}
   [\psi^i(\sigma)+i\psi^{i+1}(2\pi-\sigma)]|V_u\rangle=0,
   \quad
  i=1,\cdots, 4
\end{align}
where
$0<\sigma<\pi$,
and
the periodic boundary condition $\psi^{5}\equiv \psi^1$
is understood. 
On the other hand, 
we introduce 
the kink boundary condition by
\begin{equation}
  \label{kink cond}
  \begin{aligned}
    &[\psi^i(\sigma)+i\psi^{i+1}(2\pi-\sigma)]|V_k\rangle=0,\quad i = 1,2,3\\
    &[\psi^i(\sigma)-i\psi^{i+1}(2\pi-\sigma)]|V_k\rangle=0,\quad i = 4,
  \end{aligned}
\end{equation}
where again
$0<\sigma<\pi$.
Here, we observe that 
the boundary state -- which can be regarded as the two-vertex state -- satisfies a similar kink boundary condition:
$
[\psi^1(\sigma)+i\psi^2(2\pi-\sigma)]|B\rangle=
[\psi^2(\sigma)-i\psi^1(2\pi-\sigma)]|B\rangle=0$
($0\leq \sigma \leq \pi$).
The kink boundary condition is ``natural" from this perspective.
We will also see that 
the Neumann function method applied $p$-vertex states
with $p$ even naturally gives rise to the kind boundary condition.
We will see below that the kink boundary 
condition yields the predicted entanglement behaviors in Secs.\ \ref{sec:Corner contributions and vertex states}
and \ref{sec:reflected-entropy}.

\subsection{Direct method}
We first present the Dirac method, which is able to solve for the vertex state for both usual and kink boundary conditions. 
Since there is no interaction,
the solution to Eq.\ \eqref{usual cond} or \eqref{kink cond}
can be explicitly constructed as a coherent state
(Gaussian state).
For simplicity, we will focus below the usual boundary condition
and delegate the details for the case of 
the kink boundary condition to Appendix \ref{app:direct-kink}.
It can be numerically verified that for the kink boundary, direct calculation method gives the same result as the Neumann coefficient method. 

The boundary condition
\eqref{usual cond}
can be diagonalized by a unitary transformation.
Explicitly, we introduce the ``rotated'' fields as
$\bm{\eta}=U\bm{\psi}$ 
where the unitary matrix 
$U$ is given by
\begin{align}
U = \frac{1}{2} \left(
  \begin{array}{cccc}
    1 & 1 & 1 & 1\\
    -i & -1 & i & 1\\
    -1 & 1 & -1 & 1\\
    i & -1 & -i & 1
  \end{array}
\right).
\end{align}
The transformed fields obey the anticommutation relation
$
\{ \eta^1(\sigma),\eta^{1}(\sigma')\}
=
\{ \eta^3(\sigma),\eta^{3}(\sigma')\}
=
\{ \eta^2(\sigma),\eta^{4}(\sigma')\}
=
2\pi
\sum_{n\in \mathbb{Z}}\delta (\sigma-\sigma'-2\pi n)
$.
We note that 
the matrix $U$ diagonalizes 
the ``shift'' matrix as
\begin{equation}
\begin{aligned}
    &\left(
    \begin{array}{cccc}
        0 & 1 & 0 & 0\\
        0 & 0 & 1 & 0\\
        0 & 0 & 0 & 1\\
        1 & 0 & 0 & 0
    \end{array}
    \right)=U^\dagger
    \left(
    \begin{array}{cccc}
        1 & 0 & 0 & 0\\
        0 & i & 0 & 0\\
        0 & 0 & -1 & 0\\
        0 & 0 & 0 & -i
    \end{array}
    \right)U.
\end{aligned}
\end{equation}

The original boundary condition is translated
into the boundary condition of
the transformed fields $\eta$:
\begin{equation}
    \begin{aligned}
      &[\eta^1(\sigma)+i\mathrm{sgn}\,(\sigma)\eta^1(2\pi-\sigma)]|V_u\rangle=0,
      \\
    &[\eta^3(\sigma)-i\mathrm{sgn}\,(\sigma)\eta^3(2\pi-\sigma)]|V_u\rangle=0,
      \\
      &[\eta^2(\sigma)-\eta^2(2\pi-\sigma)]|V_u\rangle=0,
      \\
    &[\eta^4(\sigma)+\eta^4(2\pi-\sigma)]|V_u\rangle=0.
    \end{aligned}
\end{equation}
We note that the boundary conditions for $\eta^1$ and $\eta^3$
are decoupled and take the form
$  
[\eta(\sigma)-i\mathrm{sgn}\,(\sigma) e^{i\mathrm{sgn}\,(\sigma)\theta}\eta(2\pi-\sigma)]|V\rangle=0
$
($0\leq \sigma \leq 2\pi$)
where $\theta=0$ or $\pi$.
The solution to this boundary condition is derived
in \cite{2022PhRvB.105k5107L}
and given by
\begin{equation}
  |V\rangle \propto \exp
  \Big(
  \sum_{r,s\geq 1/2}
  \frac{1}{2}
  K_{rs}(\theta) \eta_{-r}\eta_{-s}
  \Big)|0\rangle,
\end{equation}
where $|0\rangle$ is the ground state of the $\eta$-fermion field. 
Given $\theta$, the explicit form of $K(\theta)$ is summarized in Appendix \ref{app:K-theta}.
As for the boundary conditions for $\eta^2$ and $\eta^4$,
they can be written in terms of the Fourier modes as 
\begin{equation}
  \left[
     \left(
    \begin{array}{c}
         \eta_r^4 \\
         \eta_r^2  
    \end{array}
    \right)+\left(
    \begin{array}{cc}
        0 & 1 \\
        -1 & 0
    \end{array}
    \right) \left(
    \begin{array}{c}
         \eta_{-r}^2 \\
         \eta_{-r}^4  
    \end{array}
  \right)
\right]
|V_u\rangle =0.
\end{equation}
The modes with different $r$ decouple, which allows a simple boundary state solution. 

To summarize, the vertex state
in the $\eta$ basis
can be constructed as
$
|V_u\rangle\propto
\exp{(\sum_{r,s\geq 1/2}\frac{1}{2}\eta_{-r}^a K_{\eta, rs}^{ab}\eta_{-s}^b)}|0\rangle
$
where the matrix $K_{\eta}$ is given by
\begin{equation}
    K_\eta = \left(
    \begin{array}{cccc}
        K(\pi) & 0 & 0 & 0 \\
        0 &   0 & 0 & \bo\\
        0 & 0 & K(0) & 0 \\
        0 & -\bo & 0 & 0
    \end{array}
    \right),
\end{equation}
in the basis $(\eta_{-r}^1,\eta_{-r}^2,\eta_{-r}^3,\eta_{-r}^4)^T$.
Finally, $|V_u\rangle$
can be written 
in the original $\psi$ basis by ``rotating back'', 
\begin{equation}
  \label{Vu}
  |V_u\rangle\propto \exp{
    \Big(\sum_{r,s\geq 1/2}\frac{1}{2}\psi_{-r}^i
    K_{rs}^{ij}\psi_{-s}^j \Big)}|0\rangle,
\end{equation}
where the matrix $K$ is obtained from $K_{\eta}$
as
$
  K = U^T K_\eta U
$.

\subsection{Neumann function method}

In the Neumann function method,
we start from the Gaussian ansatz solution 
of the form \eqref{Vu}
and find the matrix $K$ such that
the vertex state reproduces the two-point correlation function
(Neumann function) on the plane.
(The latter condition can be thought of as the definition
of vertex states -- see Ref.\ \cite{2022PhRvB.105k5107L}.)
Specifically, 
we consider
the two-point correlation function of the fermion field
$\sim 1/(w-w')$ on the complex plane
and consider 
\begin{equation}
  \label{neu func}
  K^{ij}(\sigma,\sigma')=\left(\frac{\partial w_i}{i\partial \sigma}\right)^{\frac{1}{2}}\frac{1}{w_i(\sigma)-w_j(\sigma')} \left(\frac{\partial w_j}{i\partial \sigma'}\right)^{\frac{1}{2}}.
\end{equation}
Here, $w_i$ is the conformal transformation from
$p$ copies of half cylinders to the full complex plane:
\begin{equation}
  w_i(z)=w_{i,0}\Big(\frac{1+z}{1-z}\Big)^{\frac{2}{p}},\quad z=e^{i\sigma}
\end{equation}
with the constant term satisfies $w_{i+1,0}=(e^{i\pi})^{\frac{2}{p}}w_{i,0}$.
From the Fourier transform of
the Neumann function 
we read off $K^{ij}_{rs}$ as
\begin{equation}
  \label{fourier trsf}
  K^{ij}(\sigma,\sigma')=\sum_{r,s\geq 1/2}e^{ir\sigma}e^{is\sigma'}K_{rs}^{ij}
  +\delta^{ij}\sum_{r\geq 1/2}e^{-ir(\sigma-\sigma')}.
\end{equation}
From the given Neumann function \eqref{neu func}
one can check explicitly
the presence of the second term (``singular term'')
-- see Appendix \ref{app:explicit} for details.

In order to see the boundary condition
satisfied by the so-constructed vertex state,
we need to check the behavior of the Neumann function
under the reflection $\sigma\to 2\pi -\sigma$.
This amounts to $z\ra 1/z$ and leads to:
\begin{equation}
\begin{aligned}
  w_i\Big(\frac{1}{z}\Big)
  &=
    w_{i,0}
    \Big(\frac{1+\frac{1}{z}}{1-\frac{1}{z}}
    \Big)^{\frac{2}{p}}
    =w_{i,0}\Big(\frac{1+z}{1-z}\Big)^{\frac{2}{p}}
    \frac{1}{(-1)^{\frac{2}{p}}}\\
    &=w_i\frac{1}{(-1)^{\frac{2}{p}}}=w_{i-1}(z). 
\end{aligned}
\end{equation}
On the other hand,
the derivative is transformed as 
\begin{align}
  \Big(\frac{\partial w_i}{i\partial \sigma}
  \Big)^{\frac{1}{2}}\ra
  \Big(\frac{-4z}{p(1-z^2)}\frac{w_i}{(-1)^{\frac{2}{p}}}
  \Big)^{\frac{1}{2}}.
\end{align}
where we noted
\begin{align}
  \frac{\partial w_i}{i\partial \sigma}
  =w_{i,0}\frac{2}{p}
  \Big(\frac{1+z}{1-z}\Big)=\frac{4z w_i}{p(1-z^2)}.
\end{align}
To satisfy the usual boundary condition
$
K^{ij}(2\pi-\sigma,\sigma')=
-iK^{i-1,j}(\sigma,\sigma')
$
($i=1,\cdots,4$),
we need to choose to branch cut of $\omega_{i,0}^{1/2}$ such that:
\begin{equation}
  \left(\frac{w_{i,0}}{(-1)^{\frac{2}{p}-1}}\right)^{\frac{1}{2}}
  =-iw_{p-1,0}^{\frac{1}{2}}.
\end{equation}
For $p=4$, this leads to $e^{3i\pi/4}w_{i,0}^{1/2}=w_{i-1,0}^{1/2}$,
which cannot be satisfied no matter how the branch cut is chosen. 
On the other hand,
the kink boundary condition is
\begin{equation}
    \begin{aligned}
      K^{ij}(2\pi-\sigma,\sigma')&=-iK^{i-1,j}(\sigma,\sigma'),\quad i=2,3,4,\\
      K^{ij}(2\pi-\sigma,\sigma')&=iK^{i-1,j}(\sigma,\sigma'),\quad i=1,
    \end{aligned}
\end{equation}
which can be satisfied by choosing $w_{1,0}=i,w_{2,0}=-1,w_{3,0}=-i,w_{4,0}=1$ and
$w_{1,0}^{1/2}=e^{i\pi/4},w_{2,0}^{1/2}=e^{i3\pi/2},w_{3,0}^{1/2}=e^{i3\pi/4},w_{4,0}^{1/2}=1$.


To summarize, from the Neumann function
\eqref{neu func},
we can construct the vertex state $|V_k\rangle$
obeying the kink boundary condition
explicitly as a fermionic Gaussian state
with the coefficient $K^{ij}_{rs}$ in \eqref{fourier trsf}. 
More details and the explicit form of the
matrix $K^{ij}_{rs}$ are given in Appendix \ref{app:explicit}. 

\begin{table}[t]
    \centering
    \begin{tabular}{c|c|c|c|c}
    \hline
    \hline
    \multicolumn{2}{c|}{Partition} & 1 & 2 & 3\\    
    \hline
    \multicolumn{2}{c|}{} &\includegraphics[width=0.2\linewidth]{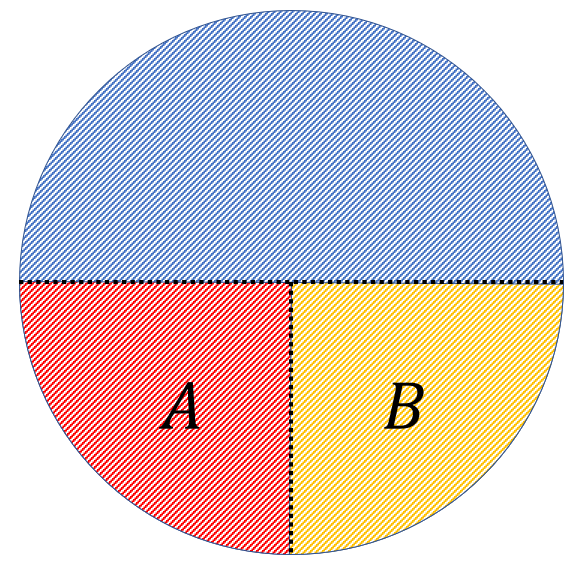}& \includegraphics[width=0.2\linewidth]{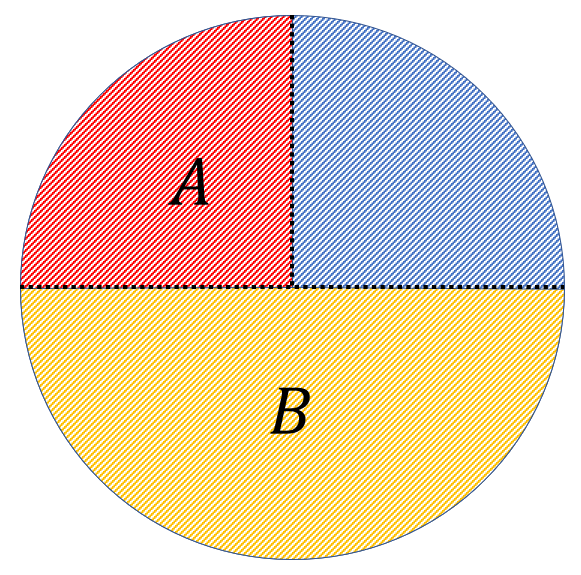} & \includegraphics[width=0.2\linewidth]{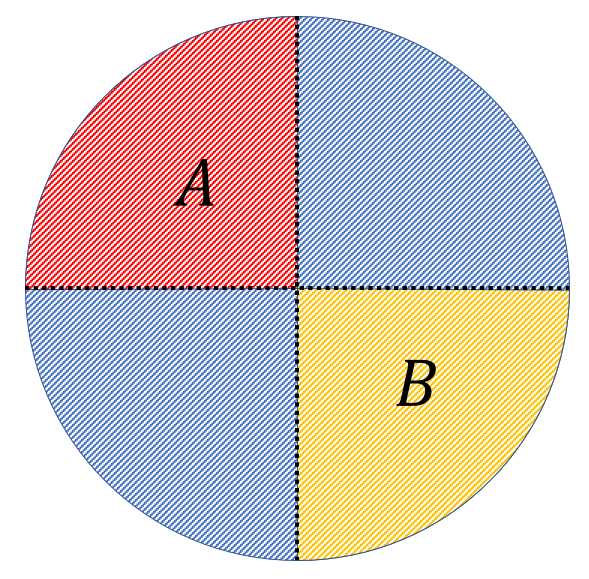}  \\
      \hline
                                   & (p) & 41.2389 & 41.1811 & 82.2467\\
      $S_{AB}$ &(k)  & 41.2389& 41.1811  &  82.2467\\
                                   & (u)  & 41.2389  & 41.1811& 82.2467 \\
      \hline
                                   & (p) & 41.1234 & 41.2389 & 0.1155\\
      $I(A,B)$ & (k)   & 41.1233 & 41.2389& 0.1155\\
                                   &(u)   & 41.1233 & 41.2389 & 0.1155  \\
   \hline
      & (p) & 41.2389 & 41.3544 & 0.2471 \\ 
     $S_R(A,B)$ &(k)   & 41.2389& 41.3544 & 0.2471\\
       &(u)  &41.2276  & 41.3481  & 0.2471  \\
      \hline
                                   & (p) & $0.1155(=\frac{c}{3}\ln 2)$ & $0.1155(=\frac{c}{3}\ln 2)$ & 0.1316\\
      $h$ & (k) & 0.1155 & 0.1155 & 0.1316\\
                                   &(u)  & 0.1043& 0.1093 & 0.1316\\
      \hline
       &  (p) & 30.8425 & 30.9292 & - \\
      $\mathcal{E}(A,B)$& (k)   & 30.8185& 30.9054 & 0.0708\\ 
      &(u)  & 30.8191& 30.9054 & 0.0708\\
    \hline
    \hline
    \end{tabular}
    \caption{Three different partitions and the corresponding entanglement measures for $|V_u\rangle$ and $|V_k\rangle$ at $\beta=0.01$. For each entanglement measure, the first row (p), second row (k) and third row (u) record predicted value, numerical result using kink boundary condition, and numerical result using usual boundary condition, respectively. Reference value: for $\beta=0.01$,  $\frac{\pi^2}{24\beta}=41.1234$ and $\frac{\pi^2}{32\beta}=30.8425$.}
    \label{tab:four-string-tbl}
\end{table}

\subsection{Correlation measures}

With the explicit forms of the vertex states
$|V_u\rangle$ and $|V_k\rangle$,
we are ready to calculate the correlation measures,
i.e., entanglement entropy,
mutual information, reflected entropy, and
entanglement negativity.
As the constructed vertex states are Gaussian,
we can calculate these quantities numerically and efficiently
\cite{Peschel_2003,
Bueno_2020,
Shapourian_2017,Shapourian_2019}.
Specifically, we partition
the 4 copies of the CFTs (edge states)
into three parties, $A$, $B$, and the compliment of $A\cup B$. 
The particular partitions we consider
are listed in Table \ref{tab:four-string-tbl}. The first two partitions correspond to the case where subregions $A$ and $B$ are adjacent which can be directly compared with the analytical results in the previous sections, as we will discuss in details below. 
We set the regulator (cutoff)
$\beta=0.01$ in this subsection.
We also recall that the circumference is set to $L=2\pi$. 

First, for the entanglement entropy of bipartition $S_{AB}$,
the two boundary conditions give the same result.
For Partitions 1 and 2,
we can also check that the numerics agrees with
the prediction Eq.\ \eqref{eqn:ent-p-vertex}.
Here, we note that
we need to restore the extra contribution $\frac{c}{3}\ln\frac{p}{2}$
discussed in Eq.\ \eqref{eq:p}
and
that the topological piece is zero,
$S_{\mr{topo}}=0$,
in the free fermion model, 
\begin{align}
  S_{AB}
  &= \frac{ \pi c L}{24\beta}
  +\frac{c}{3}\ln \sin
  \frac{\pi p_{AB}}{4}
  +
  \frac{c}{3}\ln 2,
\end{align}
where $p_{AB}=2, 3$ for Partition 1 and 2, respectively.

For Partition 3 where the region $A\cup B$
consists of two disconnected parts ($A$ and $B$), 
the calculation 
in Sec.\ \ref{sec:Corner contributions and vertex states}
does not apply since 
the subregion was assumed to be simply connected there.
Nevertheless, 
recall that
we observed 
below Eq.\ \eqref{eqn:ent-p-vertex} 
the 
corner contribution is identical to
the bipartite entanglement entropy of the ground state of (1+1)-dimensional CFT. 
Motivated by this,
it is then tempting to compare the numerical result for Partition 3
with the entanglement entropy of the ground state of (1+1)-dimensional CFT 
with disjoint intervals.
For the free fermion CFT, it is given by \cite{2009JPhA...42X4005C}
  \begin{equation}
    \begin{aligned}
      S_{1d}
      &=
      \frac{c}{3}
      \ln
      \left[
        \frac{\sin |x_{21}|\sin |x_{32}|\sin |x_{43}|\sin |x_{41}|}
        {\sin |x_{31}|\sin |x_{42}|}
      \right]
      \\
            &=-\frac{2c}{3}\ln 2,
    \end{aligned}
  \end{equation}
where
the end points of the two disjoint intervals
$[x_1, x_2]$
and
$[x_3, x_4]$
using the cord coordinate are given by
$x_1=0,x_2=\pi/2,x_3=\pi,x_4=3\pi/2$,
and $x_{ij}:= x_i - x_j$.
Thus, we compare our numerics with 
\begin{align}
  \label{disjoint}
  S_{AB}=2\left(\frac{\pi c L}{24\beta}+\frac{c}{3}\ln 2\right)-\frac{2c}{3}\ln 2
  =\frac{\pi c L}{12\beta}.
\end{align}
Here, the factor of 2 in the first term comes from doubling the number of twist
operators in this case.
The mutual information $I(A,B)$ can be obtained from $S_{AB}$.
As demonstrated in Table \ref{tab:four-string-tbl},
the numerics agrees well with Eq.\ \eqref{disjoint}.

Unlike the entanglement entropy $S_{AB}$
and mutual information $I(A,B)$,
we found that the reflected entropy and Markov gap
depend on the boundary conditions. 
For Partitions 1 and 2,
we checked that the numerical results with the kind boundary condition
agree with the CFT prediction \eqref{eq:SR},
and reproduce the Markov gap $h=\frac{c}{3}\ln 2$.
For Partition 3,
the usual and kink boundary conditions 
seem to give the same result. 
Once again,
the CFT prediction \eqref{eq:SR} is not directly applicable
here since region $A$ and $B$ are not adjacent. 
Nevertheless, the numerical calculation on the 1d free fermion model shows $h_{1d}=0.1316$
(when extrapolating to $L\ra \infty$), which again agrees with the above
prediction.

We can also discuss logarithmic negativity
$\mathcal{E}(A,B)$ for the same configurations 1,2 and 3.
For the case of integer quantum Hall states,
logarithmic negativity
in these configurations 
was studied in Ref.\ 
\cite{https://doi.org/10.48550/arxiv.2208.12819}.
While we do not have the corresponding CFT calculations along the line
of Sec.\ \ref{sec:Corner contributions and vertex states}
and Sec.\ \ref{sec:reflected-entropy},
we once again borrow the corresponding one-dimensional result
\cite{Calabrese_2012},
leading to the following prediction:
\begin{align}
\mathcal{E}(A,B)=\frac{\pi c L}{32\beta}+\frac{c}{4}\ln 2+\frac{c}{4}\ln \left[
    \frac{\sin(\frac{p_A \pi}{p})\sin\big(\frac{p_B \pi}{p})}{\sin(\frac{(p_A+p_B) \pi}{p}\big)}
  \right],
\end{align}
when region $A$ and $B$ are adjacent. 
For Partition 2, the predicted value is $\frac{\pi^2}{32\beta}=30.8425$, which
does not match perfectly with 30.8185 from numerics. Nevertheless, the
difference between
the negativities for
Partition 1 and 2
is $\Delta\mathcal{E}(A,B)=0.0869$, and close to
the predicted value $\frac{c}{4}\ln 2=0.0866$. 

%
%
There is a potential ambiguity in the $O(1)$ term in the one-dimensional result for logarithmic negativity due to the OPE coefficient between twist operators. Unlike the reflected entropy, where the OPE coefficient is universal, depending only on the central charge, the OPE for negativity depends on the full operator content of the theory because the replica manifold is a Riemann surface with genus growing with replica number (see Appendix \ref{replica_OPE_app} for details). We have set the OPE coefficient to one by hand and have found good agreement with numerics.

\section{Discussion}
\label{sec:disc}

In this work, we
developed an analytical approach
to calculate the 
corner contribution to bipartite entanglement entropy
and multipartite entanglement quantities (reflected entropy and Markov gap in
particular)
for generic (2+1)-dimensional topologically-ordered ground states. 
Some of our central results are
presented in Eqs.\ \eqref{eqn:ent-p-vertex}, \eqref{eq:SR}
and \eqref{Markov gap result}.
This then supports the conjecture on the Markov gap made in the previous works
by looking at examples, 
$h= (c/3)\ln 2$.
We hope this analytical approach helps us better understand
the conjecture that relates the Markov gap and gappable boundaries. 
It is of fundamental importance to understand this better.

An important future work is to compute $h$ purely
from the bulk-perspective, using TQFT.
This was done for many entanglement quantities in various configurations 
by using surgery method in TQFT.
If this were possible, the entanglement quantity would be written
in terms of the data of the TQFT. However, TQFT knows the central charge only mod 8
(by the Gauss-Milgram formula). One would then speculate that
it would then be necessary to have, perhaps, fully-extended TQFT.

At a technical level, 
another challenge is to understand the contribution from the topological interface in the reflected entropy calculation in the general cases. In the above  calculation, we focus on the Ishibashi vacuum interface which corresponds to the $(2+1)$d ground state of topological order on the sphere. For higher genus systems, the anyon insertions in the non-contractible loops will lead to degenerate ground states $|\Psi_i\rangle$, where in general $|\Psi\rangle=\sum_i \psi_i |\Psi_i\rangle$. Whether $h$ would receive topological contribution in this general case is an open question. 
It would be also desirable to extend our calculations to incorporate the presence of
non-Abelian anyons.

\section*{Acknowledgement}
This work is supported by
the National Science Foundation under Award No.~DMR-2001181,
a Simons Investigator Grant fromthe Simons Foundation (Award No.~566116),
and 
the Gordon and Betty Moore Foundation through Grant
GBMF8685 toward the Princeton theory program.
This work was performed in part at Aspen Center for Physics, which is supported by National Science Foundation grant PHY-1607611. YL was supported in part by the National Science Foundation under Grant No. NSF PHY-1748958, the Heising-Simons Foundation, and the Simons Foundation (216179, LB) at the Kavli Institute for Theoretical Physics. JKF is supported by the Institute for Advanced Study and the National Science Foundation under Grant No.~PHY-2207584. YK is supported by the Brinson Prize Fellowship at Caltech and the U.S. Department of Energy, Office of Science, Office of High Energy Physics, under Award Number DE-SC0011632.

\begin{widetext}

\appendix
\section{Reflected entropy for pure state}
\label{app:ref-pure}

In this appendix we use conformal interface approach to reproduce a well-known result: If $\rho_{A\cup B}$ is a density matrix for a pure state, then $S_R=2S_A$. This serves as another consistency check for the validity of the conformal interface method. 

Recall that R\'enyi reflected entropy is computed using the replica trick:
\begin{equation}
    S_R^n = \lim_{m\ra 1}\frac{1}{1-n}\ln \frac{Z_{n,m}}{(Z_{1,m})^n},
\end{equation}
where $Z_{n,m}$ was defined in Eq.\ \eqref{eqn:Znm}. 
Here, $n$ is the R\'enyi replica index, and $m$ is the replica index for handling square root of the density matrix. 
In left hand side of Fig.\ \ref{fig:bi-ref}(a), we take $m=6$ as an example and draw the path integral representation for each R\'enyi replica in $Z_{n,m}$. The black dashed lines indicates the gluing within the same R\'enyi replica, while the cuts with blue and orange annotations are glued to the next (or previous) R\'enyi replica. After taking $n$ R\'enyi replicas, we obtain two cylinder path integral with circumference $4n\beta$, and $(m-2)n$ cylinder path integral with circumference $4\beta$. Note all the cylinders have length $L$ and periodic boundary condition is taken implicitly (which makes them torus). 

Similarly, the path integral representation for $(Z_{1,m})^n$ will yield $mn$ cylinder path integral with circumference $4\beta$. Combining the result of $Z_{n,m}$ and $(Z_{1,m})^n$, we find $Z_{n,m}/(Z_{1,m})^n$ is equal to the square of $Z_n/Z_1^n$ 
from the $S_A$ calculation, as shown in Fig.\ \ref{fig:bi-ref}(b). After taking the logarithmic, we come to the conclusion that $S^n_R=2 S_A^n$, thus $S_R=2S_A$ after taking the R\'enyi replica limit $n\ra 1$.


\begin{figure}
    \centering
    \includegraphics[width=0.9\textwidth]{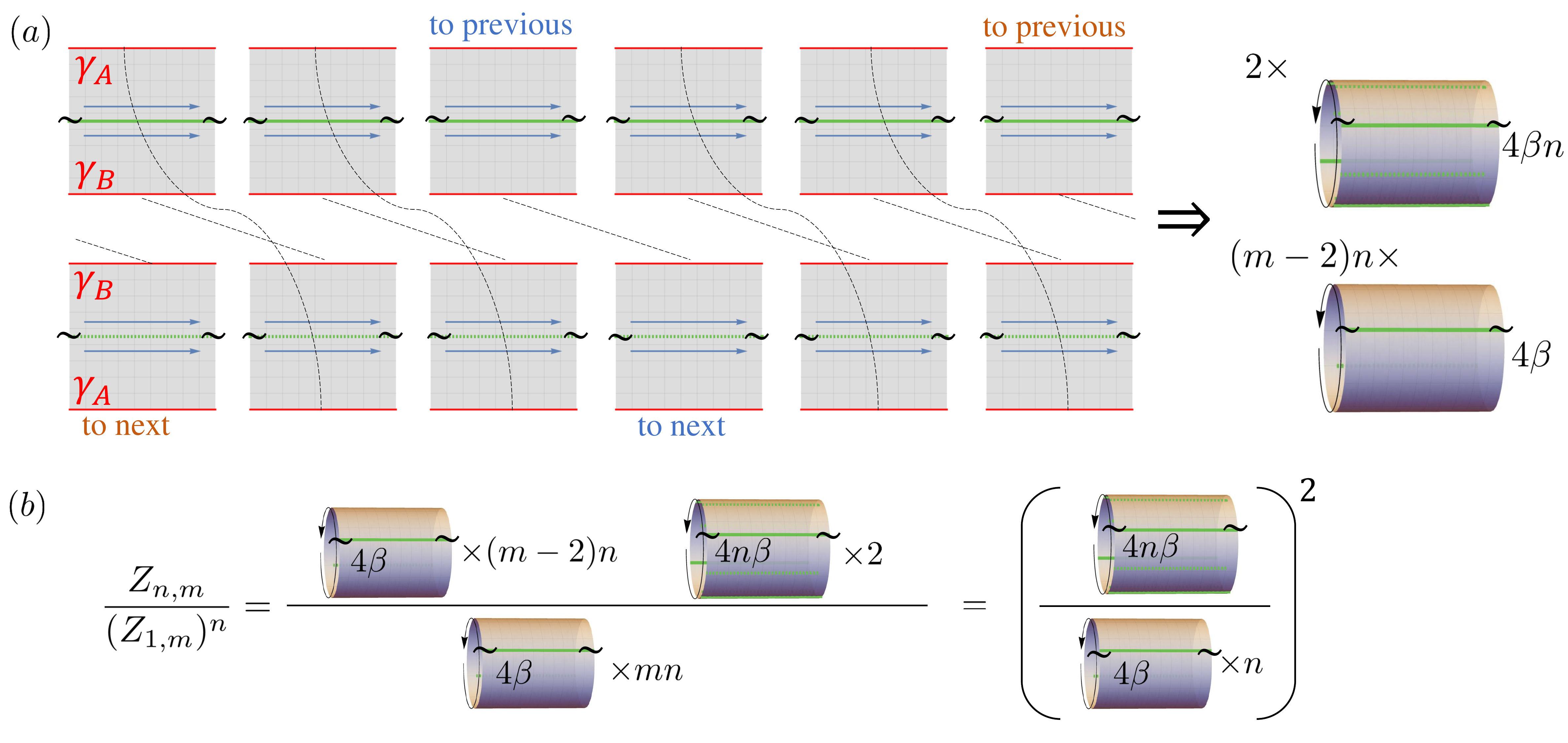}
    \caption{
   (a) Path integral representation of $Z_{n,m}$ defined in Eq.\ \eqref{eqn:Znm}, for the special case where $\rho_{A\cup B}$ is a pure density matrix. The black dashed lines indicates the gluing within 
   the same R\'enyi replica, while the cuts with blue and orange annotations are glued to the next (or previous) R\'enyi replica. The periodic boundary condition is indicated by the tilted symbol. 
   (b) Path integral representation of $Z_{n,m}/(Z_{1,m})^n$. }
    \label{fig:bi-ref}
\end{figure}

\section{Solution to single fermion boundary condition}
\label{app:K-theta}

In this section, we give the solution of a single fermion boundary condition, which is useful in the direct method for obtaining the vertex state. Here, the solution is stated without proof and we refer to \cite{2022PhRvB.105k5107L} for the detailed derivation. 

The single fermion boundary condition is formulated as:
\begin{equation}
\left[
    \eta(\sigma)+g(\sigma)\eta(2\pi-\sigma)
    \right]
    |B\rangle=0, \quad 0\leq \sigma \leq 2\pi,
    \label{eqn:single-fermion-bdy}
\end{equation}
with the consistency relation $g(\sigma)g(-\sigma)=1$. To write down the solution, let's first expand $g(\sigma)$ in terms of the Fourier modes:
$g(\sigma)=\sum_{n\in\mathbb{Z}} e^{in\sigma} g_n$. If we define matrix $N$ with components $N_{n,m}=g_{-n-m}$, the boundary condition in terms of the Fourier mode is $[\eta_r+\sum_s N_{rs} \eta_s]|B\rangle=0$, where $r,s$ runs through both positive and negative half integers. To separate the creation operators ($\eta_r$ with negative $r$) and annihilation operators ($\eta_r$ with positive $r$), we then introduce a block structure:
\begin{equation}
\begin{aligned}
  &  N = \left(
    \begin{array}{cc}
        N^{++} &  N^{+-} \\
        N^{-+} &  N^{--}
    \end{array}
    \right),\\
&N_{r,s}^{++} = N_{r,s}=g_{-r-s},\quad  N_{r,s}^{+-} = N_{r,-s}=g_{-r+s}\\
   &  N_{r,s}^{-+} = N_{-r,s}=g_{r-s},\quad 
      N_{r,s}^{--} = N_{-r-s}=g_{r+s}.
\end{aligned}
\end{equation}
In the above blocks $r,s$ only take positive values, namely,  $r,s\geq 1/2$.
We state without proof that the solution of boundary condition \eqref{eqn:single-fermion-bdy} is:
\begin{equation}
    |B\rangle \propto \exp\left(\frac{1}{2}\sum_{r,s\geq 1/2}K_{rs}\eta_{-r} \eta_{-s}\right)|0\rangle,
\end{equation}
where $K$ matrix is given in terms of the four blocks of $N$ matrix:
\begin{equation}
    K:=(\mathbbm{1}+N^{++})^{-1}(N^{+-})=(N^{-+})^{-1}(\mathbbm{1}+N^{--}).
\end{equation}

In the discussion of the vertex state, the function $g(\sigma)$ takes the form of $g(\sigma)=-i
\mathrm{sgn}\,(\sigma) e^{i
\mathrm{sgn}\,(\sigma)\theta}$, which carries a parameter $\theta$. The corresponding matrices $N$ and $K$ are denoted as $N(\theta)$ and $K(\theta)$.

\section{Direction method: kink boundary condition}
\label{app:direct-kink}

In this appendix, we solve the vertex state with kink boundary condition $|V_k\rangle$ by the direct method. As a consistency check, we have numerically verified that this solution is the same as $|V_k\rangle$ obtained from the Neumann coefficient method. 

Consider the kink boundary condition on $0<\sigma<\pi$,
\begin{equation}
    \begin{aligned}
    &[\psi^i(\sigma)+i\psi^{i+1}(2\pi-\sigma)]|V_k\rangle=0,\quad i = 1,2,3\\
    &[\psi^i(\sigma)-i\psi^{i+1}(2\pi-\sigma)]|V_k\rangle=0,\quad i = 4,
    \end{aligned}
\end{equation}
We would like to find a matrix $U$ that diagonalizes the ``shift'' matrix and use it to rotate the basis. The boundary condition thus decouples in the rotated basis.  
To start, we notice the ``shift'' matrix can be diagonalized by:
\begin{equation}
    \left(
    \begin{array}{cccc}
        0 & 1 & 0 & 0\\
        0 & 0 & 1 & 0\\
        0 & 0 & 0 & 1\\
        -1 & 0 & 0 & 0
    \end{array}
    \right)=U^\dagger  \left(
    \begin{array}{cccc}
        \frac{\sqrt{2}}{2}(1+i) & 0 & 0 & 0\\
        0 & \frac{\sqrt{2}}{2}(1-i) & 0 & 0\\
        0 & 0 & \frac{\sqrt{2}}{2}(-1-i) & 0\\
        0 & 0 & 0 & \frac{\sqrt{2}}{2}(-1+i)
    \end{array}
    \right) U,
\end{equation}
with unitary matrix $U$:
\begin{equation}
     U = \frac{1}{2} \left(
    \begin{array}{cccc}
        \frac{\sqrt{2}}{2}(-1+i) & i & \frac{\sqrt{2}}{2}(1+i) & 1\\
        \frac{\sqrt{2}}{2}(-1-i) & -i & \frac{\sqrt{2}}{2}(1-i) & 1\\
        \frac{\sqrt{2}}{2}(1-i) & i & \frac{\sqrt{2}}{2}(-1-i) & 1\\
        \frac{\sqrt{2}}{2}(1+i) & -i & \frac{\sqrt{2}}{2}(-1+i) & 1
    \end{array}
    \right).
\end{equation}
The rotated real fermions are thus $\bm{\eta}=U\bm{\psi}$. By 
\begin{equation}
UU^T =\left(
\begin{array}{cccc}
    0 & 1 & 0 & 0  \\
    1 & 0 & 0 & 0\\
    0 & 0 & 0 & 1\\
    0 & 0 & 1 & 0
\end{array}
\right),    
\end{equation}
the anticommutation relation of the rotated basis $\bm{\eta}$ is $\lbrace\eta^1_r,\eta^2_s\rbrace=\delta_{r+s,0},\lbrace\eta^3_r,\eta^4_s\rbrace=\delta_{r+s,0}$. 

Let's first consider the $\eta^1,\eta^2$ pair. For $\eta^1$, the boundary condition is:
\begin{equation}
    \begin{aligned}
    &[\eta^1(\sigma)+ie^{i\frac{\pi}{4}}\eta^1(2\pi-\sigma)]|V_{(12)}\rangle=0,\quad 0<\sigma<\pi\\
    &[\eta^1(\sigma)-ie^{-i\frac{\pi}{4}}\eta^1(2\pi-\sigma)]|V_{(12)}\rangle=0,\quad \pi<\sigma<2\pi
    \end{aligned}
\end{equation}
This amounts to choosing $\theta_1 = \frac{5}{4}\pi$ in the boundary condition $  
[\eta(\sigma)-i\mathrm{sgn}\,(\sigma) e^{i\mathrm{sgn}\,(\sigma)\theta}\eta(2\pi-\sigma)]|V\rangle=0
$
($0\leq \sigma \leq 2\pi$). Similarly, for $\eta^2$, the corresponding angle is $\theta_2 = \frac{3}{4}\pi$. Now in terms of the Fourier modes, the boundary condition can be rewritten as:
\begin{equation}
     \left[\left(
    \begin{array}{c}
         \eta_r^2 \\
         \eta_r^1  
    \end{array}
    \right)+\sum_s\left(
    \begin{array}{cc}
        0 & N_{rs}(\theta_2=\frac{3\pi}{4}) \\
        N_{rs}(\theta_1=\frac{5\pi}{4}) & 0
    \end{array}
    \right) \left(
    \begin{array}{c}
         \eta_{s}^1 \\
         \eta_{s}^2  
    \end{array}
\right)\right]|V_{(12)}\rangle=0.
\end{equation}
We can separate the creation and annihilation parts explicitly, where $r,s>0$:
\begin{equation}
    \left[\left(
    \begin{array}{c}
         \eta_r^2 \\
         \eta_r^1\\
         \eta_{-r}^1\\
         \eta_{-r}^2
    \end{array}
    \right)+
    \sum_{s\geq 1/2}\left(
    \begin{array}{cccc}
        N_{rs}^{++}(\theta_2) & 0 & 0 & N_{rs}^{+-}(\theta_2)  \\
        0 & N_{rs}^{++}(\theta_1) & N_{rs}^{+-}(\theta_1) & 0\\
        0 & N_{rs}^{-+} (\theta_1) & N_{rs}^{--}(\theta_1) & 0\\
        N_{rs}^{-+}(\theta_2) & 0 & 0 & N_{rs}^{--}(\theta_2)
    \end{array}
    \right)\left(
    \begin{array}{c}
         \eta_s^2 \\
         \eta_s^1\\
         \eta_{-s}^1\\
         \eta_{-s}^2
    \end{array}
    \right)\right]|V\rangle=0,
\end{equation}
from which we read out:
\begin{equation}
\begin{aligned}
    N_{(12)}^{++}&=\left(
    \begin{array}{cc}
        N^{++}(\theta_2) & 0 \\
        0 & N^{++}(\theta_1)
    \end{array}
    \right),\quad 
    N_{(12)}^{+-}=\left(
    \begin{array}{cc}
        0 & N^{+-}(\theta_2) \\
        N^{+-}(\theta_1) & 0
    \end{array}
    \right),\\
     N^{-+}_{(12)}&=\left(
    \begin{array}{cc}
        0 & N^{-+}(\theta_1) \\
        N^{-+}(\theta_2) & 0
    \end{array}
    \right),\quad
     N^{--}_{(12)}=\left(
    \begin{array}{cc}
        N^{--}(\theta_1) & 0 \\
        0 & N^{--}(\theta_2)
    \end{array}
    \right).
\end{aligned}
\end{equation}
Using the four block matrices, the vertex state solution for $\eta^1,\eta^2$ pair is:
\begin{equation}
\begin{aligned}
    &|V_{(12)}\rangle\propto \exp\left(\frac{1}{2}\sum_{r,s\geq\frac{1}{2}}\sum_{i,j=1,2}\eta_{-r}^i K_{(12),rs}^{ij} \eta_{-r}^j\right)|0\rangle,\\
    &\mathrm{with}\quad K_{(12)}=(\bo+N_{(12)}^{++})^{-1}(N_{(12)}^{-+}).
\end{aligned}
\end{equation}

Similarly, for the $\eta^3,\eta^4$ pair, the corresponding angles are $\theta_3=\frac{\pi}{4}$ and $\theta_4=-\frac{\pi}{4}$, and $K_{(34)}$ can be obtained in a similar way. Finally, we need to rotate back to the original $\psi$ basis. The vertex state solution in the  basis $\psi$ is thus:
\begin{equation}
\begin{aligned}
& |V_k\rangle \propto \exp\left(\frac{1}{2}\sum_{r,s\geq\frac{1}{2}}\sum_{i,j=1,2,3,4}\bm{\psi}^i_{-r} K^{ij}_{rs} \psi_{-s}^j\right)|0\rangle,\\
 &   \mathrm{with}\quad K=U^T 
    \left(
    \begin{array}{cc}
        K_{(12)} & 0 \\
        0 &  K_{(34)}
    \end{array}
    \right)
    U.
\end{aligned}
\end{equation}

Although the vertex state solution for kink boundary condition $|V_k\rangle$ can be obtained from either the direct method or the Neumann coefficient method, the Neumann coefficient method is more desirable for numerical calculation, because it does not involve the matrix inverse which would lead to numerical inaccuracy.

\section{Neumann coefficient method: Explicit form of solution}
\label{app:explicit}
In this appendix, we write down the explicit form of vertex state solution $|V_k\rangle$ with kink boundary condition by finding the mode expansion of $K^{ij}(z,z')$ in Eq.\ \eqref{neu func}:
\begin{equation}
    K^{ij}(\sigma,\sigma')=\left(\frac{\partial w_i}{i\partial \sigma}\right)^{\frac{1}{2}}\frac{1}{w_i(\sigma)-w_j(\sigma')} \left(\frac{\partial w_j}{i\partial \sigma'}\right)^{\frac{1}{2}}.
\end{equation}
Here $w$ is the conformal transformation that brings $p$ copies of half cylinders to a complex plane:
\begin{equation}
    w_i(z)=w_{i,0}(\frac{1+z}{1-z})^{\frac{1}{2}},\quad z=e^{i\sigma},
\end{equation}
with the constant terms satisfy $w_{i+1,0}=i w_{i,0}$. In the following we aim to write $K^{ij}(\sigma,\sigma')$ in terms of mode expansion $z^{n+\frac{1}{2}}, (z')^{m+\frac{1}{2}}$:
\begin{equation}
    K^{ij}(\sigma,\sigma')=\sum_{n,m\geq 0} K^{ij}_{nm} z^{n+\frac{1}{2}} (z')^{m+\frac{1}{2}}.
\end{equation}
To do so, we will write down the mode expansion of $(\frac{\partial w_i}{i\partial \sigma})^{1/2}$ and $\frac{1}{w_i(\sigma)-w_j(\sigma')}$. 

Let us start from the mode expansion of $(\frac{\partial w_i}{i\partial \sigma})^{1/2}$. 
Using:
\begin{equation}
    (\frac{\partial w_i}{i\partial \sigma})=w_{i,0} \left(\frac{1-z}{1+z}\right)^{1/2}\frac{z}{(1-z)^2},
\end{equation}
and we denote
\begin{equation}
    g(z)=(\frac{1+z}{1-z})^{\frac{1}{4}},
\end{equation}
the factor $(\frac{\partial w_i}{i\partial \sigma})^{1/2}$ can be rewritten as:
\begin{equation}
    (\frac{\partial w_i}{i\partial \sigma})^{1/2}=(w_{i,0})^{1/2}\frac{z^{1/2}}{1-z}g(-z).
    \label{eqn:Neumann-1}
\end{equation}
We thus need to obtain the mode expansion of $g(z)$. 

For $g(z)=(\frac{1+z}{1-z})^{\frac{1}{4}}=\sum_{n\geq 0} g_n z^n$, the coefficients $g_n$ satisfy the recursion relation:
\begin{equation}
    \frac{1}{2}g_n=(n+1) g_{n+1}-(n-1)g_{n-1},
\end{equation}
and the first two coefficients are $g_0=1, g_1=\frac{1}{2}$. This with the recursion relation allows us to obtain all the $g_n$. 

Next, we would need to rewrite $ \frac{1}{w_i(\sigma)-w_j(\sigma')}$. By using $\frac{1}{m-n}=\frac{m^3+m^2 n+m n^2+n^3}{m^4-n^4}$ we can obtain:
\begin{equation}
\begin{aligned}
    \frac{1}{w_i(\sigma)-w_j(\sigma')}=&\frac{(1-z)^{\frac{1}{2}} (1-z')^{\frac{1}{2}}}{4(z-z')(1-zz')}\left\lbrace w_{i,0}^3(1+z)^{\frac{3}{2}}(1-z')^{\frac{3}{2}}+w_{i,0}^2 w_{j,0}(1+z)(1-z')(1+z')^{\frac{1}{2}}(1-z)^{\frac{1}{2}}\right.\\
    &\left.+w_{i,0}w_{j,0}^2(1+z)^{\frac{1}{2}}(1-z')^{\frac{1}{2}}(1+z')(1-z)+w_{j,0}^3(1+z')^{\frac{3}{2}}(1-z)^{\frac{3}{2}}\right\rbrace.
\end{aligned}
\label{eqn:Neumann-2}
\end{equation}

Combining \eqref{eqn:Neumann-1} and \eqref{eqn:Neumann-2}, $K^{ij}(z,z')$ is rewritten as:
\begin{equation}
    \begin{aligned}
        K^{ij}(z,z')=&\left[(w_{i,0})^{\frac{1}{2}}(w_{j,0})^{\frac{1}{2}}\frac{z^{\frac{1}{2}}(z')^{\frac{1}{2}}}{4(z-z')(1-zz')}\right]\frac{g(-z)g(-z')}{(1-z)^{\frac{1}{2}}(1-z')^{\frac{1}{2}}}\\
       & 
       \times \left\lbrace w_{i,0}^3(1+z)^{\frac{3}{2}}(1-z')^{\frac{3}{2}}+w_{i,0}^2 w_{j,0}(1+z)(1-z')(1+z')^{\frac{1}{2}}(1-z)^{\frac{1}{2}}\right.\\
    &\left.
    \quad +w_{i,0}w_{j,0}^2(1+z)^{\frac{1}{2}}(1-z')^{\frac{1}{2}}(1+z')(1-z)+w_{j,0}^3(1+z')^{\frac{3}{2}}(1-z)^{\frac{3}{2}}\right\rbrace.
    \end{aligned}
\end{equation}
There are four terms in the curly bracket. 

For the four-vertex state, $i,j$ take values from $\lbrace 1,2,3,4\rbrace$. Since $K^{ij}$ only depends on the difference between $i$ and $j$, there are four difference circumstances. 
For $i=j$:
\begin{equation}
    K^{ii}=\frac{z^{\frac{1}{2}} (z')^{\frac{1}{2}}}{2(z-z')}[g(z)g(-z')+g(-z)g(z')].
    \label{eqn:K1}
\end{equation}
For $j=i+1$:
\begin{equation}
    K^{i,i+1}=\frac{z^{1/2}(z')^{1/2}}{2(1-zz')}\left(\frac{w_{i,0}}{w_{j,0}}\right)^{-\frac{1}{2}}\left[g(z)g(-z')+ig(-z)g(z')\right].
\end{equation}
For $j=i+2$:
\begin{equation}
    K^{i,i+2}=\frac{z^{\frac{1}{2}}(z')^{\frac{1}{2}}}{2(z-z')} \left(\frac{w_{i,0}}{w_{j,0}}\right)^{-\frac{1}{2}}[g(z)g(-z')-g(-z)g(z')].
\end{equation}
For $j=i+3$:
\begin{equation}
    K^{i,i+3}=\frac{z^{1/2}(z')^{1/2}}{2(1-zz')}\left(\frac{w_{i,0}}{w_{j,0}}\right)^{-\frac{1}{2}}\left[g(z)g(-z')-ig(-z)g(z')\right].
    \label{eqn:K4}
\end{equation}

To evaluate the expansion coefficients of $K^{ij}$, we need to use the following two relations:
\begin{equation}
    \begin{aligned}
        &\frac{g(z)g(-z')}{2(z-z')}=\sum_{n,m\geq 0}P_{nm}^+ z^n (z')^m+\mathrm{sing.}\\
        &\mathrm{with}\quad P_{nm}^+=\frac{1}{n+m+1}[g_{n+1}g_{m+1}(n+1)(m+1)-g_n g_m nm](-1)^m,
    \end{aligned}
    \label{eqn:relation-1}
\end{equation}
where $\mathrm{sing.}$ denotes the singular term $\frac{1}{2(z-z')}$; 
and
\begin{equation}
    \begin{aligned}
        &\frac{g(z)g(-z')}{2(1-zz')}=\sum_{n,m\geq 0, n\neq m} P_{nm}^- z^n (z')^m+\sum_{m\geq 0} P_m z^m(z')^m \\
        &\mathrm{with}\quad P_{nm}^-  =
        \begin{cases}
        \frac{1}{n-m}[n g_n (m+1)g_{m+1}-g_{n+1}(n+1)g_m m](-1)^m & n\neq m\\
        \frac{1}{2}\sum_{0\leq n\leq m}g_n^2(-1)^n & n=m
        \end{cases}.
    \end{aligned}
    \label{eqn:relation-2}
\end{equation}
The above two relations can be derived using:
\[
(\partial_\rho+\partial_{\rho'}+1)\frac{g(z)g(-z')}{z-z'}=-2(1+zz')\partial \partial'(g(z)g(-z'))
\]
and
\[
(\partial_\rho-\partial_{\rho'})\frac{g(z)g(-z')}{1-zz'}=-2(z+z')\partial \partial' (g(z)g(-z')).
\]
 
By plugging \eqref{eqn:relation-1} and \eqref{eqn:relation-2} into \eqref{eqn:K1} - \eqref{eqn:K4}, we obtain the mode expansion coefficients of $K^{ij}(z,z')$. We summarize the explicit expressions below.

\subsection{Summary} To summarize, the mode expansions for $K^{ij}$ are:
\begin{equation}
\begin{aligned}
    K^{ii}&=\sum_{n\geq 0,m\geq 0}z^{n+\frac{1}{2}}(z')^{m+\frac{1}{2}} (P_{nm}^+-P_{mn}^+) +\sum_{n\geq 0}(\frac{z'}{z})^{n+\frac{1}{2}}\\
    K^{i,i+1}&=\left(\frac{w_{i,0}}{w_{i+1,0}}\right)^{-\frac{1}{2}}\left[ \sum_{n\geq 0,m\geq 0} z^{n+\frac{1}{2}}(z')^{m+\frac{1}{2}}(P_{nm}^- +i P_{mn}^-)  \right] \\
    K^{i,i+2} & = \left(\frac{w_{i,0}}{w_{i+2,0}}\right)^{-\frac{1}{2}}\sum_{n\geq 0,m\geq 0}z^{n+\frac{1}{2}}(z')^{m+\frac{1}{2}} (P_{nm}^+ +P_{mn}^+)\\
    K^{i,i+3}& = \left(\frac{w_{i,0}}{w_{i+3,0}}\right)^{-\frac{1}{2}}\left[ \sum_{n\geq 0,m\geq 0} z^{n+\frac{1}{2}}(z')^{m+\frac{1}{2}}(P_{nm}^- -i P_{mn}^-) \right].
\end{aligned}
\end{equation}
One nice feature is that the singular terms in $K^{ii}$ take the form:
\begin{equation}
   \frac{z^{\frac{1}{2}}(z')^{\frac{1}{2}}}{z-z'}=\sum_{n\geq 0}(\frac{z'}{z})^{n+\frac{1}{2}},
\end{equation}
which is what we desired in order to satisfy the boundary condition (as discussed near Eq.\ \eqref{fourier trsf}). 

Let's denote $K_0=P^+ - (P^+)^T$, $K_1=P^-+i(P^- )^T$, $K_2=P^+ + (P^+)^T$ and  $K_3=P^- - i(P^-)^T$. 
The whole $K$ matrix is:
\begin{equation}
    K = \left(
    \begin{array}{cccc}
        K_0 & \left(\frac{w_{1,0}}{w_{2,0}}\right)^{-\frac{1}{2}} K_1
         & \left(\frac{w_{1,0}}{w_{3,0}}\right)^{-\frac{1}{2}}K_2 & \left(\frac{w_{1,0}}{w_{4,0}}\right)^{-\frac{1}{2}} K_3\\
 \left(\frac{w_{2,0}}{w_{1,0}}\right)^{-\frac{1}{2}} K_3 & K_0 & \left(\frac{w_{2,0}}{w_{3,0}}\right)^{-\frac{1}{2}} K_1 & \left(\frac{w_{2,0}}{w_{4,0}}\right)^{-\frac{1}{2}} K_2\\
    \left(\frac{w_{3,0}}{w_{1,0}}\right)^{-\frac{1}{2}} K_2 & \left(\frac{w_{3,0}}{w_{2,0}}\right)^{-\frac{1}{2}} K_3 & K_0 & \left(\frac{w_{3,0}}{w_{4,0}}\right)^{-\frac{1}{2}} K_1\\
    \left(\frac{w_{4,0}}{w_{1,0}}\right)^{-\frac{1}{2}} K_1 & \left(\frac{w_{4,0}}{w_{2,0}}\right)^{-\frac{1}{2}} K_2 & \left(\frac{w_{4,0}}{w_{3,0}}\right)^{-\frac{1}{2}} K_3 & K_0
    \end{array}
    \right).
\end{equation}

As a consistency check, 
let us verify $K$ is antisymmetric. Note that by definition $K_0$ is antisymmetric and $K_2$ is symmetric. Using our previous choice of branch cut ($w_{1,0}=i,w_{2,0}=-1,w_{3,0}=-i,w_{4,0}=1$ and $w_{1,0}^{1/2}=e^{i\pi/4},w_{2,0}^{1/2}=e^{i3\pi/2},w_{3,0}^{1/2}=e^{i3\pi/4},w_{4,0}^{1/2}=1$), the factors are $(\frac{w_{1,0}}{w_{3,0}})^{-1/2}=i$, $(\frac{w_{2,0}}{w_{4,0}})^{-1/2}=i$, so the blocks associated with $K_2$ has the desired property under tranposition. For the blocks associated with $K_1$ and $K_3$, for example:
\begin{equation}
    \left(
    \begin{array}{cc}
        0 &  \left(\frac{w_{1,0}}{w_{2,0}}\right)^{-\frac{1}{2}} K_1\\
         \left(\frac{w_{2,0}}{w_{1,0}}\right)^{-\frac{1}{2}} K_3 & 0
    \end{array}
    \right)=\left(
    \begin{array}{cc}
        0 &  \left(\frac{w_{1,0}}{w_{2,0}}\right)^{-\frac{1}{2}} (P^-+i(P^-)^T)\\
         \left(\frac{w_{2,0}}{w_{1,0}}\right)^{-\frac{1}{2}} (P^--i(P^-)^T) & 0
    \end{array}
    \right). 
\end{equation}
Using our previous choice, $(\frac{w_{1,0}}{w_{2,0}})^{-1/2}=e^{i5\pi/4}\sim -1-i$. One can check:
\[
\begin{aligned}
&(-1-i)(P^-+i(P^-)^T)^T=(-1-i)((P^-)^T+iP^-),\\
&(-1+i)(P^- -i(P^-)^T)=(1+i)(P^-)^T + (-1+i)P^-,
\end{aligned}
\]
which shows this block is indeed antisymmetric. Note that the antisymmetric property is dependent on the choice of $w_0$. If we make another choice of branch cut $(\frac{w_{1,0}}{w_{2,0}})^{-1/2}=e^{i3\pi/4}$, this block would not be antisymmetric.

\section{One-dimensional CFT}
\label{replica_OPE_app}
In 1+1D CFT, one may use the twist operator formalism to compute various quantum information inequalities. In this appendix, we review how to evaluate the mutual information, reflected entropy, and logarithmic negativity for adjacent intervals in the vacuum.

The moments of the reduced density matrix are given by the two-point function of twist operators located at the entangling surfaces \cite{2009JPhA...42X4005C}
\begin{align}
    \text{Tr}\, \rho_A^n = \langle \sigma_n(x_1) \bar{\sigma}_n(x_2)\rangle ,
\end{align}
which is fixed by conformal symmetry
\begin{align}
    \text{Tr}\, \rho_A^n = |x_1 - x_2|^{-4h_n}.
\end{align}
Taking the replica limit, we find
\begin{align}
    S_{vN}(A) = \frac{c}{3}\ln\frac{x_2 - x_1}{\epsilon}.
\end{align}
Because the union of adjacent intervals is again an interval, we immediately find
\begin{align}
    I(A,B) = \frac{c}{3}\ln
    \frac{(x_2 - x_1)(x_3-x_2)}{\epsilon(x_3-x_1)}.
\end{align}

The reflected entropy for adjacent intervals may also be computed using the twist operator formalism as a three-point function \cite{2021JHEP...03..178D}
\begin{align}
    S_R(A,B) = \lim_{n,m\rightarrow 1}\frac{1}{1-n} \ln \langle \sigma_{g_A}(x_1) \sigma_{g_{A}^{-1} g_B}(x_2)\sigma_{g_B^{-1}}(x_3) \rangle.
\end{align}
This is also completely fixed by conformal symmetry, up to a nontrivial OPE coefficient
\begin{align}
    S_R(A,B) = \frac{c}{3}\ln
    \frac{(x_2 - x_1)(x_3-x_2)}{\epsilon(x_3-x_1)}+ \lim_{n,m\rightarrow 1}\frac{1}{1-n} \ln
    C_{n,m} .
\end{align}
The evaluation of this OPE coefficient is crucial because it is precisely what leads to a nontrivial Markov gap. It is useful to consult the Riemann-Hurwitz formula to compute the genus of the replica manifold
\begin{align}
    g = \frac{1}{2}\sum r_j -s + 1,
\end{align}
where $s$ is the total number of replicas and $r_j+1$ is the number of sheets that meet at the $j^{th}$ branch point. From the definition of the twist fields, one concludes that the genus for all $n$ and $m$ is zero, so the OPE coefficient is \textit{universal}, only dependent on the central charge of the theory. Explicitly, it is given by \cite{2021JHEP...03..178D}
\begin{align}
    C_{n,m} = (2m)^{-4h_n},
\end{align}
which precisely leads to the universal $\frac{c}{3}\ln 2$ Markov gap.

We proceed to the negativity, which is computed by a different three-point correlation function of twist operators \cite{Calabrese_2012}
\begin{align}
    \mathcal{E}(A,B) = \lim_{n_e\rightarrow 1} \ln
    \langle \sigma_{n_e}(x_1) \bar{\sigma}^{(2)}_{n_e}(x_2)\sigma
    _{n_e}(x_3) \rangle,
\end{align}
where the middle twist operator applies a double cyclic permutation. The limit is taken from the even integers to one. Of course, this is still fixed by conformal symmetry, up to the OPE coefficient
\begin{align}
    \mathcal{E}(A,B) = \frac{c}{4}\ln \frac{(x_2 - x_1)(x_3-x_2)}{\epsilon(x_3-x_1)}+\lim_{n_e\rightarrow 1} \ln C_{n_e},
\end{align}
with the same scaling as mutual information and reflected entropy. A significant different arises however when considering the OPE coefficient. The replica manifold for even $n_e$ can be seen to have genus $\frac{n_e}{2}-1$. Partitiln functions on non-zero genus surfaces depend on the full operator content of the theory so the OPE coefficient is nonuniversal. We therefore cannot expect a simple analytic function in $n_e$ analogous to the reflected entropy. In the main text, we have set $C_1 = 1$ by hand, finding good agreement with numerics.

\end{widetext}
\bibliography{main}

\end{document}